\documentclass[microtype]{jloganal}

\usepackage{pinlabel}
\usepackage{latexsym}
\usepackage[all]{xy}

\title[A Coinductive Approach to Computing with Compact Sets]{A Coinductive Approach to Computing with Compact Sets}

\author[Ulrich Berger]{Ulrich Berger}
\givenname{Ulrich}
\surname{Berger}
\address{Department of Computer Science\\
Swansea University\\\newline
Swansea SA2 8PP, United Kingdom}
\email{U.Berger@swansea.ac.uk}
\urladdr{http://www-compsci.swan.ac.uk/~csulrich/} 

\author[Dieter Spreen]{Dieter Spreen}
\givenname{Dieter}
\surname{Spreen}
\address{Department of Mathematics\\
University of Siegen\\\newline 
57068 Siegen, Germany\\ and\\\newline
 Department of Decision Sciences\\
University of South Africa\\\newline 
Pretoria 0003, South Africa}
\email{spreen@math.uni-siegen.de}
\urladdr{http://www.uni-siegen.de/fb6/tcs/team/spreen/}


\arxivreference{cs.LO/1510.08498}
\arxivpassword{xuet5}

\volumenumber{}
\issuenumber{}
\publicationyear{}
\papernumber{}
\startpage{}
\endpage{}
\doi{}
\MR{}
\Zbl{}
\received{}
\revised{}
\accepted{}
\published{}
\publishedonline{}
\proposed{}
\seconded{}
\corresponding{}
\editor{}
\version{}

\newtheorem{claim}{\it Claim}
\theoremstyle{plain}
\newtheorem{theorem}{Theorem}[section]
\newtheorem*{sdthm}{Soundness Theorem}
\newtheorem{lemma}[theorem]{Lemma}
\newtheorem{proposition}[theorem]{Proposition}
\newtheorem{corollary}[theorem]{Corollary}

\theoremstyle{definition}
\newtheorem{definition}[theorem]{Definition}

\newcommand{\II}{\mathbb{I}}
\newcommand{\NN}{\mathbb{N}}

\newcommand{\QQ}{\mathbb{Q}}

\newcommand{\RR}{\mathbb{R}}

\newcommand{\PPP}{\mathcal{P}}
\newcommand{\QQQ}{\mathcal{Q}}
\newcommand{\TTT}{\mathcal{T}}
\newcommand{\AV}{\mathrm{AV}}
\newcommand{\av}[1]{\mathrm{av}_{#1}}
\newcommand{\ball}[3]{\mathrm{B}_{#1}(#2,#3)}
\newcommand{\cball}[3]{\mathrm{B}^\le_{#1}(#2,#3)}
\newcommand{\coco}{\mathrm{C}}
\newcommand{\compact}{\mathrm{K}}
\newcommand{\hdm}{\mu_{\mathrm H}}
\newcommand{\treehdm}{\delta_{\mathrm H}}
\newcommand{\hdr}{\eta_\mathrm{H}}

\newcommand{\mds}[1]{\mathcal{#1}}
\newcommand{\SD}{\mathrm{SD}}
\newcommand{\val}[1]{[\![#1]\!]}
\newcommand{\set}[2]{\mbox{$\{\,#1 \mid #2 \,\}$}}
\newcommand{\fun}[3]{\mbox{$#1 \colon #2 \rightarrow #3$}}
\newcommand{\pfun}[3]{\mbox{$#1 \colon #2 \rightharpoonup #3$}}
\newcommand{\pair}[1]{\langle #1 \rangle}
\newcommand{\card}[1]{\mathopen\parallel #1 \mathclose\parallel}
\def\dom{\mathop{\mathstrut\rm dom}}
\def\range{\mathop{\mathstrut\rm range}}
\def\int{\mathop{\mathstrut\rm int}\nolimits}
\newcommand{\cantor}{\mathbf{C}}
\newcommand{\fminus}{f_{-}}
\newcommand{\fplus}{f_{+}}

\newcommand{\rea}[2]{#1\,\mathbf{r}\,#2}
\newcommand{\powplus}{\mathcal{P}_{+}}
\newcommand{\branches}[1]{E_{#1}}
\newcommand{\subtree}[2]{#1_{#2}}
\newcommand{\powfin}{\mathcal{P}^+_{\mathrm{fin}}}
\newcommand{\cauchy}{\mathrm{A}}
\newcommand{\basic}{\mathrm{K}_0}
\newcommand{\prea}[1]{\mathbf{R}(#1)}
\newcommand{\reap}[1]{\widetilde{#1}}
\newcommand{\pleft}[1]{\mathbf{p}_0(#1)}
\newcommand{\pright}[1]{\mathbf{p}_1(#1)}
\newcommand{\pow}[1]{\mathcal{P}(#1)}
\newcommand{\powf}[1]{\mathcal{P}_{\mathrm{fin}}}

\begin{document}

\begin{abstract}
  Exact representations of real numbers such as the signed digit
  representation or more generally linear fractional representations or the
  infinite Gray code represent real numbers as infinite streams of digits. In
  earlier work by the first author it was shown how to extract certified
  algorithms working with the signed digit representations from constructive
  proofs. In this paper we lay the foundation for doing a similar thing with
  nonempty compact sets. It turns out that a representation by \emph{streams}
  of finitely many digits is impossible and instead \emph{trees} are needed.
\end{abstract}

\begin{asciiabstract}
Exact representations of real numbers such as the signed digit
  representation or more generally linear fractional representations or the
  infinite Gray code represent real numbers as infinite streams of digits. In
  earlier work by the first author it was shown how to extract certified
  algorithms working with the signed digit representations from constructive
  proofs. In this paper we lay the foundation for doing a similar thing with
  nonempty compact sets. It turns out that a representation by streams
  of finitely many digits is impossible and instead trees are needed.
  \end{asciiabstract}
  
  \maketitle
  

\section{Introduction}
\label{sec-intro}
Digital representations of real numbers have been widely studied in the 
literature. Probably best known is the signed digit representation as 
considered in 
Ciaffaglione and Di Gian\-antonio \cite{CiaffaglioneGianantonio06}, 
Escard\'{o} and Marcel-Romero \cite{EscardoMarcial-Romero07}
Berger and Hou~\cite{BergerHou08}
and many others, 
where a real number in $[-1,1]$ is represented by a stream of 
signed digits $-1,0,1$, a digit $d$ representing the mapping 
$x \mapsto (x+d)/2$. This has been generalized
to linear fractional representations studied in Edalat and 
S\"underhauf \cite{EdalatSuenderhauf98} as well as Edalat and 
Heckmann  \cite{EdalatHeckmann02}
where digits represent maps of the form $x \mapsto (ax+b)/(cx+d)$.
A variant of the signed digit representation is the infinite
Gray code introduced by Tsuiki \cite{Tsuiki02} which represents real 
numbers in $[-1,1]$ by a binary stream
with possibly one undefined entry. The infinite Gray code has the 
remarkable property
that it is non-redundant, that is, every real has exactly one code.

Berger \cite{Berger11} showed how to use the method of program extraction 
from proofs 
(see eg Schwichtenberg and Wainer \cite{SchwichtenbergWainer12})
to extract certified algorithms 
working with the signed digit representations. 
In this paper we lay the foundation for doing a similar
thing with nonempty compact sets. 

In the first part of the paper 
(\fullref{sec-digit-space} to \fullref{sec-computable}) 
we develop our approach in the style of classical mathematics 
speaking explicitly about representations. 
In \fullref{sec-digit-space} we generalize the signed digit representation to 
\emph{digit spaces} $(X,D)$, that is, we replace the interval $[-1,1]$ 
by a complete bounded metric space $X$ and the signed digits by a 
finite set $D$ of contractions whose images cover $X$.
In \fullref{sec-compact-hausdorff} and \fullref{sec-tree} we introduce the hyper-space
of nonempty compact sets and develop a representation of compact sets by non-wellfounded 
\emph{trees} of digits, after having shown that in most situations a representation by
\emph{streams} of digits is impossible.
In \fullref{sec-computable} we compare the notion of computability for compact sets 
generated by our tree representation with those studied by Brattka and Presser~\cite{BrattkaPresser03}.

In the second part
(\fullref{sec-coind} to \fullref{sec-equiv}) we recast the approach constructively 
in a representation free way.
We introduce a coinductive predicate on the powerset of $X$ whose realizers 
are trees representing nonempty compact subsets of $X$ generalizing the
coinductive approach to the signed digit representation studied by Berger \cite{Berger11}.
We sketch how this approach can be used to extract programs computing
with compact sets from constructive proofs and comment on the relation to 
iterated function systems as studied by Edalat \cite{Edalat97}. The Cantor set is considered as an example.

\section{Digit spaces}
\label{sec-digit-space}
A metric space $X$ with metric $\mu$ is called \emph{bounded} if there exists
a number $M\ge 0$, called \emph{bound} of $X$, such that
$\mu(x,y)\le M$ for all $x,y\in X$.  
A \emph{contraction} on $X$ is a function $d:X\to X$ such that there exists 
a number $q<1$, called \emph{contraction factor}, with  
$\mu(d(x),d(y))\le q\cdot\mu(x,y)$ for all $x,y\in X$. 
\begin{definition}
\label{def-digit-space}
A \emph{digit space} $(X,D)$ consists of
a bounded and complete nonempty metric space $X$ and a 
finite set $D$ of contractions on $X$, called \emph{digits}, 
that cover $X$, that is, $\bigcup\{\,d[X] \mid d\in D\,\} = X$, where $d[X] := \set{d(x)}{x \in X}$. 
\end{definition}
Our running example of a digit space will be the \emph{signed digit space}
$(\II,\AV)$ where $\II := [-1,1] \subseteq \RR$ with the 
usual metric and $\AV := \{\,\av{i} \mid i\in\SD\,\}$ with 
$\SD := \{-1,0,1\}$ and $\av{i}(x) := (x+i)/2$.
By a slight abuse of notation we will call both the elements of $\AV$ and of 
$\SD$ \emph{signed digits} or just \emph{digits}.

In the following we identify a finite sequence of digits 
$\vec d = [d_0,\ldots,d_{n-1}]\in D^n$ with the composition
$d_0\circ\cdots\circ d_{n-1}$ and a digit $d$ with the singleton sequence $[d] \in D^1$. The set of all finite sequences of digits will be denoted by $D^{< \omega}$.
Moreover, we let $q < 1$ be a common contraction factor for all digits $d \in D$.

\begin{lemma}
\label{lem-digit-compact}
Every digit space is compact, that is, the underlying metric space is
compact.
\end{lemma}
\begin{proof}
Let $\mds{D} = (X,D)$ be a digit space. 
It suffices to show that $X$ is totally bounded. So let $\varepsilon>0$.
Let $n\in\NN$ be so large that $q^n\cdot M < \varepsilon$ where $q$ is a common
contraction factor for each $d\in D$. Then for 
arbitrarily chosen $x\in X$, the set $\{\,\vec d(x) \mid \vec d\in D^n\,\}$ 
is a finite $\varepsilon$-covering, ie the $\card{D^n}$ many balls $\ball{\mu}{\vec d (x)}{\varepsilon}$ of radius $\varepsilon$ around $\vec{d}(x)$ cover $X$. 
\end{proof}
Conversely, every compact metric space is bounded, since the metric is continuous. Note that for any $x_0 \in X$, the set of all elements $\vec{d}(x_0)$ with $\vec{d} \in D^{< \omega}$ is dense in $X$.

The purpose of a digit space is to provide representations for the
elements of $X$ by infinite streams of digits. 
Let $D^\omega$ be the set of all infinite sequences of elements of $D$ 
and set for  $\alpha \in D^\omega$:
$$\alpha^{<n} := [\alpha_0,\ldots,\alpha_{n-1}]$$ 
\begin{lemma}
\label{lem-digit-stream}
Let $(X,D)$ be a digit space. Then
$\bigcap_{n \in \NN} \alpha^{< n}[X]$
is a singleton for every $\alpha \in D^\omega$. 
\end{lemma}
\begin{proof}
Since digits are continuous and $X$ is compact, the set $\alpha^{<n}[X]$ 
is compact as well. As $X$ is Hausdorff, it is closed in particular. 
Moreover, it is nonempty, since $X$ is nonempty. 
Since clearly $\alpha^{< n+1}[X] \subseteq \alpha^{< n}[X]$,
the family $(\alpha^{< n}[X])_{n \in \NN}$ has the finite 
intersection property. 
Therefore, $\bigcap_{n \in \NN} \alpha^{< n}$ is not empty. 
On the other hand, 
\[
\sup \{\, \mu(x,y) \mid x, y \in \alpha^{< n}[X] \,\} \le q^n \cdot M,
\]
which implies that 
$\sup \{\, \mu(x,y) \mid x, y \in \alpha^{< n}[X] \,\} \rightarrow 0$, 
as $n \rightarrow \infty$. 
Consequently, $||\bigcap_{n \in \NN} \alpha^{< n}[X] || \le 1$. 
\end{proof}
\begin{definition}
\label{def-digit-stream-value}
Let $(X,D)$ be a digit space and $\alpha \in D^\omega$.
In view of Lemma~\ref{lem-digit-stream}, we let $\val{\alpha}\in X$ 
denote the uniquely determined element of 
$\bigcap_{n \in \NN} \alpha^{< n}[X]$.
\end{definition}

Note that for every infinite sequence $\alpha_0\alpha_1\ldots$ of digits and every digit $d$, $d\alpha_0\alpha_1\ldots$ is an infinite sequence of digits as well.

\begin{lemma}\label{lem-valrec}
For every $d \in D$ and $\alpha \in D^\omega$, $\val{d\alpha} = d(\val{\alpha})$.
\end{lemma}
\begin{proof}
We have that $d[\{\val{\alpha}\}]  \subseteq d[\bigcap_{n\in\NN} \alpha^{<n}[X]] \subseteq d[\alpha^{<m}[X]]$, for all $m \in \NN$, and hence that $\{d(\val{\alpha})\} \subseteq \bigcap_{n\in\NN}(d\alpha)^{<n}[X] =\{\val{d\alpha}\}$. Consequently, $\val{d\alpha} = d(\val{\alpha})$.
\end{proof}

The next technical lemma will be useful in what follows.

\begin{lemma}\label{lem-gensur}
Let $\vec{d} \in D^n$. Then, for every $x \in \vec{d}[X]$, there is some $\alpha\in D^\omega$ with $\alpha^{< n} = \vec{d}$ and $\val{\alpha} = x$.
\end{lemma}
\begin{proof}
Let $x \in \vec{d}[X]$ and let $y\in X$ such that $x= \vec{d}(y)$. 
By the covering property of $D$, there exist $e \in D$ and $z\in X$ such that $y=e(z)$.
Hence $x=(\vec{d} e)(z)$, in particular $x \in (\vec{d}e)[X]$. 
By dependent choice, there is
some $\alpha \in D^\omega$ with $\alpha^{< n} = \vec{d}$ and 
$x \in \alpha^{< m}[X]$ for all $m\ge n$, hence $\val{\alpha} = x$.
\end{proof} 

As is well known, 
$D^\omega$ is a compact bounded metric space with metric
\[
\delta(\alpha, \beta) := 
\begin{cases}
0& \text{ if $\alpha = \beta$,}\\
2^{- \min \set{n}{\alpha_n\neq \beta_n}}& \text{ otherwise.}
\end{cases}
\]

\begin{proposition}\label{prop-valcont}
\begin{enumerate}
\item \label{valcont1} $\fun{\val{\cdot}}{D^\omega}{X}$ is onto and uniformly continuous.  
\item \label{valcont2} The metric topology in $X$ is equivalent to the quotient topology induced by $\val{\cdot}$.
\end{enumerate}
\end{proposition}
\begin{proof}
(\ref{valcont1}) As a consequence of the preceding lemma, $\val{\cdot}$ is onto. For the verification of the remaining statement let $\alpha, \beta$ be distinct elements of $D^\omega$. Then $$\val{\alpha}, \val{\beta} \in \alpha^{< \min \set{n}{\alpha_n \neq \beta_n}}[X],$$ and therefore 
\begin{equation}\label{eq-mudelta}
\mu(\val{\alpha}, \val{\beta}) \le q^{\min \set{n}{\alpha_n \neq \beta_n}} \cdot M.
\end{equation}
By definition of $\delta(\alpha, \beta)$, $\min\set{n}{\alpha_n \neq \beta_n} = - \log_2 \delta(\alpha. \beta)$. It follows that
\begin{equation}\label{eq-unicont}
\begin{split}
\mu(\val{\alpha}, \val{\beta}) &\le q^{- \log_2 \delta(\alpha, \beta)} \cdot M \\
&= 2^{- \log_2 q \cdot \log_2 \delta(\alpha, \beta)} \cdot M = \delta(\alpha, \beta)^{- \log_2 q} \cdot M,
\end{split}
\end{equation}
from which we obtain that $\val{\cdot}$ is uniformly continuous.

(\ref{valcont2}) We have to show that for any subset $O$ of $X$, $O$ is open in the metric topology if, and only if, $\val{\cdot}^{-1}[O]$ is open in the metric topology on $D^\omega$. 

The `only-if'-part holds as $\val{\cdot}$ is continuous. For the other direction assume that $\val{\cdot}^{-1}[O]$ is open in $D^\omega$ and $x \in O$. We need to show that $x$ lies in the interior of $O$.

Since $x \in O$, there is some $\alpha \in D^\omega$ so that $\val{\alpha} = x$. Hence, $\alpha \in \val{\cdot}^{-1}[O]$. Since the latter set is open, there is some $m_\alpha \in \NN$ with $\ball{\delta}{\alpha}{2^{-m_\alpha}} \subseteq \val{\cdot}^{-1}[O]$, which means that $\val{\beta} \in O$, for all $\beta \in D^\omega$ with $\delta(\alpha, \beta) < 2^{-m_\alpha}$.

\begin{claim}
\label{cl-qtop-1}
$\exists m \in \NN\, \forall \alpha, \beta \in D^\omega \,[\delta(\alpha, \beta) < 2^{-m} \wedge \val{\alpha} = x \Rightarrow \val{\beta} \in O]$

In words, there exists a natural number $m$ such that for all sequences $\alpha$ and $\beta$ in $D^\omega$ the first $m$ elements of which coincide, if $\val{\alpha} = x$ then $\val{\beta} \in O$.
\end{claim}

Since $\val{\cdot}$ is continuous, we have that $\val{\cdot}^{-1}[\{ x \}]$ is closed and hence compact. Moreover,
\[
\val{\cdot}^{-1}[\{ x \}] \subseteq \bigcup \set{\ball{\delta}{\alpha}{2^{-{m_\alpha}+1}}}{\val{\alpha} = x}.
\]
Hence, there exist $\alpha_1, \ldots, \alpha_n \in D^\omega$ with $\val{\alpha_i} = x$, for $1 \le i \le n$, so that
\[
\val{\cdot}^{-1}[\{ x \}] \subseteq \bigcup \set{\ball{\delta}{\alpha_i}{2^{-{m_{\alpha_i}}+1}}}{1 \le i \le n}.
\]
Set $m := \max \set{m_{\alpha_i}}{1 \le i \le n} + 1$ and let $\alpha, \beta \in D^\omega$ with $\val{\alpha} = x$. Then there is some $1 \le i \le n$ with $\delta(\alpha, \alpha_i) < 2^{-m_{\alpha_i}+1}$. It follows that
\[
\delta(\alpha_i, \beta) \le \delta(\alpha_i, \alpha) + \delta(\alpha, \beta) < 2^{-m_{\alpha_i}+1} + 2^{-m} \le 2^{-m_{\alpha_i}}.
\]
Hence, $\val{\beta} \in O$. Thus, Claim~\ref{cl-qtop-1} is proven.

Let $m$ be as in Claim~\ref{cl-qtop-1} and set 
$
E := \set{\vec d \in D^{m+1}}{\vec d [X] \subseteq O}
$
as well as 
\[ A := \bigcup\set{\vec e \,[X]}{\vec e \in E}\;\text{and}\;
   B := \bigcup\set{\vec e \,[X]}{\vec e \in D^{m+1}\setminus E}.
\]
Then $A$ and $B$ are closed sets (since they are finite unions
of compact, hence closed sets) that cover $X$. Furthermore, $A\subseteq O$.
It follows that $X\setminus B$ is an open subset of $O$. Therefore, it suffices
to show that $x\in X\setminus B$.
Assume to the contrary that $x\in B$. Then there is some $\vec e\in D^{m+1}$
such that $x \in \vec{e}\,[X]$ and there exists some 
$y \in \vec{e}\,[X] \setminus O$. 
With Lemma~\ref{lem-gensur} it moreover follows that there are 
$\alpha, \beta \in D^\omega$ with prefix $\vec e$ such that 
$\val{\alpha} = x$ and $\val{\beta} = y$. 
Thus $\delta(\alpha, \beta) < 2^{-m}$. With Claim~\ref{cl-qtop-1} 
we therefore obtain  that $y \in O$, a contradiction.
\end{proof}

Note that the proof of Proposition~\ref{prop-valcont}(\ref{valcont1}) shows that 
$\val{\cdot}$ is even H\"older continuous of order $-\log_2 q$. 
In particular if $q\le 1/2$, then $\val{\cdot}$ is Lipschitz-continuous 
with Lipschitz constant $M$.

Continuous images of compact sets are compact again. The next result is therefore a consequence of Proposition~\ref{prop-valcont}.
\begin{corollary}\label{cor-cant-x}
Let $C \subseteq D^\omega$ be compact. Then $\val{C}$ is a compact subset of $X$.
\end{corollary}

Conversely, if $A \subseteq X$ is compact, then $A$ is closed, as $X$ is Hausdorff. Hence, $C := \val{\cdot}^{-1}[A]$ is closed as well. Since $D^\omega$ is compact, it follows that $C$ is compact too. Moreover, $\val{C} = A$, because $\val{\cdot}$ is onto.
\begin{lemma}\label{lem-x-cant}
Let $A$ be a compact subset of  $X$. Then there is a compact subset $C$ of $D^\omega$ with $\val{C} = A$.
\end{lemma}

\section{Compact sets and the Hausdorff metric}
\label{sec-compact-hausdorff}

We start this section by deriving some facts about metric digit spaces 
needed in the sequel.

\begin{definition}
Let $(X, \mu)$ be a metric space and $\varepsilon \in \mathbb{R}_+$. 
Then a subset $E$ of $X$ is an \emph{$\varepsilon$--chain} if 
$\mu(x,y) \ge \varepsilon$, for all $x, y \in E$ such that $x \not= y$.
\end{definition}

\begin{lemma}\label{lem-cbnd}
Let $(X, D)$ be a digit space such that $(X, \mu)$ is bounded with bound $M$. 
Moreover, let $q$ be a uniform contraction factor for all digits in $D$. 
Then the size of every $q^n \cdot M$--chain in $X$ is bounded by $\card{D}^n$. 
\end{lemma}
\begin{proof}
Let $E$ be a $q^n \cdot M$--chain in $X$. Because of the covering property 
we have that $X = \bigcup\{\, \vec{d}[X] \mid \vec{d} \in D^n \,\}$. Now, 
let $x, y \in E$ with $x \not= y$ and assume that there is some 
$\vec{d} \in D^n$ so that $x, y \in \vec{d}[X]$. Then there are $x', y' \in X$ 
with $x = \vec{d}(x')$ and $y = \vec{d}(y')$. Hence
\begin{equation}\label{eq-bnd}
 \mu(x,y) \le q^n \cdot \mu(x', y').
\end{equation}

Since $M$ is a bound of $X$, we have that $\mu(x', y') \le M$. On the other 
hand, $\mu(x,y) \ge q^n \cdot M$, as $E$ is a $q^n \cdot M$--chain. With 
(\ref{eq-bnd}) it thus follows that $q^n \cdot M < q^n \cdot M$, a 
contradiction. Hence, $\card{E \cap \vec{d}[X]} \le 1$, for each 
$\vec{d} \in D^n$. Thus, $\card{E} \le \card{D}^n$. 
\end{proof}

For a metric space $(X,\mu)$ we denote by $\compact(X)$ the set of 
nonempty compact subsets of $X$. The \emph{Hausdorff metric}, $\hdm$, 
on $\compact(X)$ is defined by the formula
\[\hdm(A,B) := \inf\{\,\varepsilon\ge 0\mid A \subseteq\ball{\mu}{B}{\varepsilon} \land
                                      B \subseteq\ball{\mu}{A}{\varepsilon} \,\}, \]
where 
$\ball{\mu}{A}{\varepsilon} := \{\,x\in X\mid\exists y\in A\,\mu(x,y)<\varepsilon\,\}$. 

The subsequent properties are often useful. Let to this end
\[
\mu(A, y) := \inf\nolimits_{x \in A} \mu(x,y) \quad \text{and} \quad \mu_A(B) := \sup\nolimits_{y \in B} \mu(A,y).
\]
\begin{lemma}\label{lem-hausm}
For $x, y \in X$, $A, A', B, B' \in \compact(X)$ and a contraction $\fun{d}{X}{X}$ the following statements hold:
\begin{enumerate}

\item\label{lem-hausm-1} $\hdm(\{ x \}, \{ y \}) = \mu(x, y)$

\item\label{lem-hausm-2} $\hdm(A,B) = \max \{ \mu_A(B), \mu_B(A) \}$

\item\label{lem-hausm-3} $\hdm(A \cup A', B \cup B') \le \max\{ \hdm(A, B), \hdm(A', B') \}$

\item\label{lem-hausm-4} $\hdm(d[A], d[B]) \le q \cdot \hdm(A, B)$

\end{enumerate}
\end{lemma}

Note that $\compact(X)$ has the same bound $M$ as $X$. Moreover, it is well-known that $\compact(X)$ inherits completeness and compactness from $X$.
However, we cannot expect $\compact(X)$ to
have a finite covering system of contractions as we show in the 
following.

\begin{lemma}\label{lem-cptchain}
Let $(X, \mu)$ be a bounded metric space and $\varepsilon \in \mathbb{R}_+$. 
If $X$ has an $\varepsilon$--chain $E$ of size $m$, then $K(X)$ has an 
$\varepsilon$--chain of size $2^m - 1$.
\end{lemma}
\begin{proof}
It suffices to show that the collection of all nonempty subsets of $E$ is 
an $\varepsilon$--chain with respect to the Hausdorff metric. Let $A, B$ be 
two different nonempty subsets of $E$. We have to show that 
$\mu_H(A,B) \ge \varepsilon$. Since $A \not= B$, $A \setminus B$ or 
$B \setminus A$ are not empty. Without restriction we consider the first 
case. Let $x \in A \setminus B$ and $z \in B$. Then $x \not= z$ and hence 
$\mu(x,z) \ge \varepsilon$. 
It follows that $\mu_H(A,B) \ge \varepsilon$. 
\end{proof}
\begin{corollary}\label{cor-lin}
Let $ (X, \mu)$ be a bounded metric space and $\compact(X)$ have a 
finite set of 
digits. Then there are constants $c_1, c_2 \in \mathbb{R}$ such that for 
every $2^{-n}$--chain $E$ in $X$, $\card{E} \le c_1 \cdot n + c_2$.
\end{corollary}
\proof
Let $M$ be a bound of $X$, $D$ be the set of digits of $\compact(X)$ 
of size $k$ and
$q$ be a uniform contraction factor of the digits in $D$.  
If $E'$ is a $q^m \cdot M$--chain in $X$, it follows with Lemma~\ref{lem-cptchain} that $\compact(X)$ 
has a $q^m \cdot M$--chain of size $2^{\card{E'}} -1$.  By Lemma~\ref{lem-cbnd} we therefore have that
 $2^{\card{E'}} \le k^m + 1 \le k^{m+1}$. Thus,
\[
\card{E'} \le \log_2 (k^{m+1}) = (m+1) \cdot \log_2 k.
\]
Now, let $E$ be a $2^{-n}$--chain in $X$ and $m = \lfloor \frac{n+ \log_2 M}{-\log_2 q} \rfloor$. Then
\[
q^m \cdot M \le 
q^{\frac{n+ \log_2 M}{-\log_2 q}}\cdot M = 
(2^{\log_2 q})^{\frac{n+ \log_2 M}{-\log_2 q}} \cdot M = 
2^{-n}.
\]
It follows that $E$ is a $q^m \cdot M$--chain as well. Therefore, 
by the above:
\begin{align*}
\card{E} &\le m \cdot \log_2 k + \log_2 k \\ 
&\le \frac{\log_2 k}{-\log_2 q} \cdot (n + \log_2 M) + \log_2 k\\
&= \frac{\log_2 k}{-\log_2 q} \cdot n + 
(\log_2 k + \frac{\log_2 k}{-\log_2 q} \cdot \log_2 M)  \rlap{\hspace{6.3em} \qedsymbol}
\end{align*}

\begin{lemma}
\label{lem-convex}
Let $X$ be a non-trivial bounded and convex subset of a normed linear space.
Then there cannot be finitely many contractions on $\compact(X)$ that 
cover $\compact(X)$.
\end{lemma}
\begin{proof}
Let $x, y$ be different elements of $X$. Hence $|| x - y || \ge 2^{-n_0}$
for some $n_0 \in \mathbb{N}_+$. Let $n \in \mathbb{N}_+$ and set 
\[
x_\nu :=  2^{-n}  \nu x + (1 - 2^{-n}  \nu) y
\]
as well as $E_n := \{\, x_\nu \mid 0 \le \nu \le 2^n \,\}$. Then $E_n \subseteq X$
is a $2^{-(n_0 + n)}$--chain, since for $\nu \not= \nu'$ we 
have that:
\begin{align*}
|| x_\nu - x_{\nu'} || &= || (\nu - \nu') 2^{-n} x + (\nu' - \nu) 2^{-n} y || \\
& = |\nu - \nu'| 2^{-n} ||x - y|| \ge |\nu - \nu'| 2^{-(n_0 + n)} \ge 2^{-(n_0 + n)}
\end{align*}

Now, assume that $\compact(X)$ has a finite system of digits. 
By Corollary~\ref{cor-lin} there are constants $c_1, c_2 \in \mathbb{R}_+$ 
(independent of $n$) such that:
\[
\card{E_n} \le c_1 \cdot (n_0 + n) + c_2 = c_1 \cdot n + (c_1 \cdot n_0 + c_2)
\]
On the other hand, $\card{E_n} = 2^n + 1$, which is a contradiction for
large enough $n$.
\end{proof}

\section{Representation of compact sets by trees} 
\label{sec-tree}
Since, according to Lemma~\ref{lem-convex}, it will in most cases be 
impossible to turn $\compact(X)$ into a digit space, we consider a 
representation of compact sets by \emph{trees} (instead of streams) 
of digits of the original digit space $(X,D)$.
\begin{definition}
\label{def-tree}
Let $(X,D)$ be a digit space. A \emph{digital tree} is a nonempty set 
$T\subseteq D^{<\omega}$ of finite sequences of digits that 
is downwards closed under the prefix ordering and has no maximal element, that is,
$[]\in T$ and whenever
$[d_0,\ldots,d_{n}]\in T$, then $[d_0,\ldots,d_{n-1}]\in T$ and
$[d_0,\ldots,d_{n},d]\in T$ for some $d\in D$.
\end{definition}
Note that each such tree is finitely branching as $D$ is finite.
Moreover, every element $[d_0,\ldots,d_{n-1}]\in T$ can be continued to an 
infinite path $\alpha$ in $T$, that is, $\alpha \in D^\omega$ is such that $\alpha_i = d_i$, for $i < n$, and 
$[\alpha_0,\ldots,\alpha_{k-1}]\in T$ for all $k\in\NN$.
In the following we write  $\alpha\in T$ to mean that $\alpha$ is a 
path in $T$ and by a path we always mean an infinite path.

Let $\TTT_D$ denote the set of digital trees with digits in $D$ and for $T \in \TTT_D$ and $n \ge 0$, let $T^{\le n}$ be the finite initial subtree of $T$ of height $n$. Then: 
\[
T^{\le n} = \{\, \alpha^{<m} \mid \alpha \in T \wedge m \le n \,\}
\]
Every such initial subtree defines a map $\fun{f_{T,n}}{X}{\PPP(X)}$ from $X$ into the powerset of $X$ in the obvious way:
\[
f_{T,n}(x) := \set{\vec{d}(x)}{\vec{d} \in D^n \cap T}
\]

\begin{definition}
\label{def-val}
For every $T \in \TTT_D$ we define its \emph{value} by
\[
\val{T} := \bigcap\nolimits_{n\in\NN} f_{T,n}[X].
\]
\end{definition}

\begin{lemma}
\label{lem-semT}
$\val{T} =  \set{\val{\alpha}}{\alpha\in T}$.
\end{lemma}
\begin{proof}
Observe that $f_{T,n}[X] = \bigcup\set{\vec{d}[X]}{\vec{d} \in D^n \cap T}$. Therefore, $\val{\alpha} \in \val{T}$, for every $\alpha \in T$. 
Conversely, let $x \in \val{T}$. Then there is some $\vec{d} \in D^n\cap T$ with $x \in \vec{d}[X]$, for each $n \in \NN$. Since the set of all $\vec{d} \in D^n\cap T$ with $x \in \vec{d}[X]$ is a finitely branching infinite tree, it follows with K\"onig's Lemma that there is a path $\alpha \in T$ with $x \in \alpha^{<n}[X]$, for all $n$. Thus, $x = \val{\alpha}$.
\end{proof}

\xyoption{line}
\begin{corollary}\label{cor-tree-valrec}
Let $T \in \TTT_D$ and let $\{d_1,\ldots,d_m\}$ be the set of digits $d\in D$
such that $[d]\in T$ and $T_i := \{\,\vec d\in T^{<\omega} \mid d_i\vec d\in T\,\}$ 
($i=1,\ldots,m$),
ie $T_i$ is the $i$th  immediate subtree of $T$:
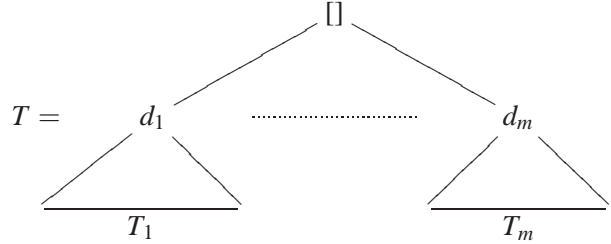
\[
\xymatrix{
&&&&[]\ar@{-}[dll]\ar@{-}[drr]&&&& \\
&T=&d_1\ar@{-}[dl]\ar@{-}[dr]&\ar@{..}@*{[thicker]}[rr]&&&d_m\ar@{-}[dl]\ar@{-}[dr]&& \\
&\ar@{-}[rr]_{\text{\normalsize $T_1$}}&&&&&&\ar@{-}[ll]^{\text{\normalsize $T_m$}}&
}
\]
Then $\val{T} = \bigcup^m_{i=1} d_i[\val{T_i}]$.
\end{corollary}
\begin{proof}
Apply Lemma~\ref{lem-valrec}.
\end{proof}
Note that this interpretation of a digital tree corresponds to the IFS-tree
of an iterated function system in Edalat \cite{Edalat97}.

Let $T$ be a digital tree and $\alpha$ be an infinite path not lying in $T$. Then there is a finite initial segment $\alpha^{< n}$ of $\alpha$ that is not contained in $T$, since as explained above, a path is identified with the sequence of its finite initial segments. Because $T$ is closed under taking initial segments, for all infinite continuations $\beta$ of $\alpha^{< n}$ we have that $\beta \not\in T$ as well. Thus, $D^\omega \setminus T$ is open in the metric topology on $D^\omega$.

\begin{lemma} \label{lem-treecl}
For any digital tree the set of its infinite paths is closed in $D^\omega$.
\end{lemma}

Obviously, the set of infinite paths of a tree is nonempty. Conversely, let $C$ be a nonempty closed subset of $D^\omega$ and $T^C := \set{\alpha^{< n}}{\alpha \in C \wedge n \in \NN}$. Obviously, $T^C$ is a digital tree.

\begin{lemma}
Let $C$ be a nonempty closed subset of $D^\omega$. Then the sequences in $C$ are exactly the paths of $T^C$.
\end{lemma}
\begin{proof}
Clearly, every element of $C$ is a path in $T^C$. Conversely, if $\beta$ is 
a path in $T^C$, then its initial segment of length $n$ is of the form 
$\alpha^{<n}$ for some $\alpha\in C$. Hence $\beta$ has distance 
$2^{-n}$ from $\alpha$. It follows that $\beta$ is in the closure of $C$,
hence in $C$.
\end{proof}

As we have already seen in Corollary~\ref{cor-cant-x} and Lemma~\ref{lem-x-cant}, the compact subsets of $X$ are exactly the images of the compact subsets in $D^\omega$ under $\val{\cdot}$. Thus, we have the following result.

\begin{lemma}
\label{lem-tree-compact}
The nonempty compact subsets of a digit space $(X,D)$ are exactly
the values of digital trees.
\end{lemma}

The metric defined on $D^\omega$ in Section~\ref{sec-digit-space} can be transferred to $\TTT_D$. As we will see next, it coincides with the Hausdorff metric.

\begin{lemma}\label{lem-treehausd}
For $S, T \in \TTT_D$,
\[
\treehdm(S,T) =
\begin{cases}
0 & \text{ if $S = T$,}\\
2^{- \min \set{n}{S^{\le n}\not= T^{\le n}}}  & \text{otherwise.}
\end{cases}
\]
\end{lemma}
\begin{proof}
Without restriction let $S \not= T$.  Then there exists 
\[
n_0 := \min \set{n}{S^{\le n} \not= T^{\le n}}.
\]
Note that $n_0 \ge 1$ as $S^{\le 0} = T^{\le 0}$. It follows that $S^{\le n_0-1} = T^{\le n_0-1}$. Let $\beta \in S$. Then there is some $\alpha \in T$ such that $\beta^{<n_0-1} = \alpha^{<n_0-1}$, ie $\min\set{n}{\alpha_n \not= \beta_n} \ge n_0$. Thus, $\delta(\alpha, \beta) \le 2^{-n_0}$ and hence $\delta(\alpha, \beta) < 2^{-n_0} + \tau$, for all $\tau > 0$. This shows that $S \subseteq \ball{\delta}{T}{2^{-n_0}+\tau}$, for all $\tau > 0$. Similarly, we obtain that $T \subseteq \ball{\delta}{S}{2^{-n_0}+\tau}$, for all $\tau > 0$. It follows that $\treehdm(S,T) \le 2^{-n_0}$.

Now, assume that $\treehdm(S,T) < 2^{-n_0}$. Then there is some $\varepsilon \in \mathbb{R}$ with $\treehdm(S,T) < \varepsilon < 2^{-n_0}$. Let $\beta \in S$. Then there exists $\alpha \in T$ with $\delta(\alpha, \beta) < \varepsilon$, from which we obtain that
\[
\min\set{n}{\alpha_n \not= \beta_n} > - \log_2 \varepsilon > n_0
\]
and hence that
\[
\min\set{n}{\alpha_n \not= \beta_n} \ge \lceil - \log_2 \varepsilon \rceil > n_0, 
\]
respectively, 
\[
\max\set{n}{\alpha_n = \beta_n} \ge \lceil - \log_2 \varepsilon \rceil -1 > n_0 -1.
\]
This shows that $S^{\le n_1 -1} \subseteq T^{\le n_1 -1}$, where $n_1 = \lceil - \log_2 \varepsilon \rceil -1$. In the same way we obtain that also $T^{\le n_1 -1} \subseteq S^{\le n_1 -1}$. Thus, $S^{\le n_1 -1} = T^{\le n_1 -1}$. Since $n_1 > n_0$, we have that $n_1 - 1 \ge n_0$. However, by definition of $n_0$, $S^{\le n_0} \not= T^{\le n_0}$,  a contradiction. Thus, $\treehdm(S,T) = 2^{-n_0}$.
\end{proof}

\begin{proposition}\label{prop-conttreeval}
\begin{enumerate}
\item\label{conttreeval1}
$\fun{\val{\cdot}}{\TTT_D}{\compact(X)}$ is onto and uniformly 
continuous.
\item\label{conttreeval2}
The topology on $\compact(X)$ induced by the Hausdorff metric is equivalent 
to the quotient topology induced by $\val{\cdot}$.
\end{enumerate}
\end{proposition}
\begin{proof}
(\ref{conttreeval1}) Ontoness is a consequence of Lemma~\ref{lem-tree-compact}. For the verification of uniform continuity let $S, T \in \TTT_D$, $\alpha \in S$ and $\beta \in T$. With Inequality~(\ref{eq-unicont}) we have that
\[
\mu(\val{S}, \val{\beta}) \le \mu(\val{\alpha}, \val{\beta}) \le \delta(\alpha, \beta)^{- \log_2 q} \cdot M \le \treehdm(S,T)^{-\log_2 q} \cdot M
\]
and hence that
\[
\mu_{\val{S}}(\val{T}) \le \treehdm(S,T)^{-\log_2q} \cdot M.
\]
Similarly, we obtain that also $\mu_{\val{T}}(\val{S}) \le \treehdm(S,T)^{-\log_2q} \cdot M$ and hence that $$\hdm(\val{S}, \val{T}) \le \treehdm(S,T)^{-\log_2q} \cdot M,$$ from which the uniform continuity of $\val{\cdot}$ follows.

(\ref{conttreeval2}) The statement follows by a straightforward adaption
of the proof of Proposition~\ref{prop-valcont}(\ref{valcont2}).
\end{proof}

\section{Computably compact sets}
\label{sec-computable}

The purpose of the present paper is to provide a logic-based approach to 
computing with continuous data. In this section we compare it with 
Weihrauch's Type-Two Theory of Effectivity~\cite{Weihrauch00}.

\begin{definition}[{\rm Brattka and Presser \cite{BrattkaPresser03}}]
\label{met-computable}
Let $(X, \mu)$ be a metric space with dense subspace $Q$, say
\[
Q = \{ u_0, u_1, \ldots \},
\]
the elements of which are called \emph{basic elements}.
Then $(X, \mu, Q)$ is called \emph{computable} if the two sets
\begin{gather*}
\set{(u, v, r) \in Q \times Q \times \QQ}{\mu(u, v) < r} \\
\set{(u, v, r) \in Q \times Q \times \QQ}{\mu(u, v) > r}
\end{gather*}
are effectively enumerable, ie the function $\lambda (u,v) \in Q^2.\, \mu(u, v)$ is computable.
\end{definition}

Note that when we say that $\set{(u, v, r) \in Q \times Q \times \QQ}{\mu(u, v) < r}$ is effectively enumerable, we mean that with respect to a canonical coding $\sigma_\QQ$ of $\QQ$, the set
\[
\set{\pair{i,j,n} \in \NN}{\mu(u_i, u_j) < \sigma_\QQ(n)}
\]
is computably enumerable. Similarly, when we say that $\lambda (u,v) \in Q^2.\, \mu(u, v)$ is computable, we mean that there is a computable function $f$ such that for given $i, j \in \NN$, $f(i,j)$ is the G\"odel number of a computable function $g$ so that $(\sigma_\QQ(g(n)))_{n \in \NN}$ is a Cauchy sequence of rationals converging to $\mu(u_i, u_j)$. In what follows we will work with  finite objects such as basic elements, finite sets of basic elements, digits or finite sequences of digits directly as in the above definition and leave it to the reader to make the statements precise, if wanted. Note that by doing so we identify a digit $d$ with the letter $d$. 

\begin{definition}
\label{def-dcomp}
Let $(X, \mu, Q), (X', \mu', Q')$ be metric spaces with countable dense subspaces $Q$ and $Q'$, respectively. A uniformly continuous map $\fun{\Phi}{X}{X'}$ is \emph{computable} if it has a computable modulus of continuity and there is a procedure $g_\Phi$ which given $u \in Q$ and $n \in \NN$ computes a basic element $v \in Q'$ with $\mu'(\Phi(u), v) < 2^{-n}$.
\end{definition}

As is easily verified, the set of all computable maps on $X$ is closed under composition.

\begin{definition}\label{def-compdigsp}
Let $(X,D)$ be a digit space such that the underlying metric space $(X, \mu)$ has a countable dense subset $Q$ with respect to which it is computable. $(X, D, Q)$ is said to be a \emph{computable digit space} if, in addition, all digits $d \in D$ are computable.
\end{definition}

Let 
\begin{gather*}
\begin{split}
&\cauchy^\text{eff} := \{\, x \in X \mid \text{there is a procedure that given $n \in \NN$}\\
&\hspace{3cm} \text{computes a basic element $u \in Q$ with $\mu(x, u) < 2^{-n}$} \,\}.
\end{split}
\end{gather*}

Besides $Q$, computable digit spaces possess other canonical dense subspaces. For $x_0 \in \cauchy^\text{eff}$ set
\[
Q_D := \set{\vec d(x_0)}{\vec d \in D^{< \omega}}.
\] 
We want to show that $Q$ and $Q_D$ are \emph{effectively equivalent} in the sense that given $u \in Q$ and $n \in \NN$ we can compute a sequence $\vec d \in D^{<\omega}$ such that $\mu(u, \vec d(x_0)) < 2^{-n}$, and that similarly there is a computable function $\fun{g}{\NN \times D^{< \omega}}{Q}$ with:
\[
\forall n \in \NN, \vec d \in D^{<\omega}  \mu(\vec{d}(x_0), g(n,\vec{d})) < 2^{-n}
\]

To do so, some more requirements have to be satisfied.

\begin{definition}
\label{def-well-covering}
A digit space $(X,D)$ is \emph{well-covering} if every element of $X$ is
contained in the interior of $d[X]$ for some $d\in D$.
\end{definition}
\begin{lemma}
\label{lem-well-covering}
Let $(X,D)$ be a well-covering digit space. Then there exists 
$\varepsilon \in \QQ_+$ such that for every $x\in X$ there exists $d\in D$ with
$\ball{\mu}{x}{\varepsilon}\subseteq d[X]$. 
\end{lemma}
\begin{proof}
Assume the contrary. Then for every $n\in\NN$ there exists $x_n\in X$ such
that $\ball{\mu}{x_n}{2^{-n}}$ is not contained in $d[X]$ for any $d\in D$.
Let $x$ be an accumulation point of the $x_n$. Then clearly, $x$ is not in the 
interior of $d[X]$ for any $d\in D$.
\end{proof}

Each $\varepsilon \in \QQ_+$ as in the preceding lemma will be called a \emph{well-covering number}. 

\begin{definition}
\label{def-deceffdense}
Let $(X, D, Q)$ be a computable digit space. We call $(X, D, Q)$
\begin{enumerate}
\item\label{def-deceffdense-1}  \emph{decidable} if for $u \in Q$, $\theta \in \QQ_+$ and $d \in D$ it can be decided whether $\ball{\mu}{u}{\theta} \subseteq d[X]$;

\item\label{def-deceffdense-2} \emph{constructively dense} if there is a procedure that given $\theta \in \QQ_+$, $d \in D$ and $u \in d[X] \cap Q$ computes a $v \in Q$ such that $\mu(u, d(v)) < \theta$.

\end{enumerate}
\end{definition}

\begin{lemma}
\label{lem-Q-QD}
Let $(X, D, Q)$ be a well-covering, decidable and constructively dense digit space. Then, for every $u \in Q$ and $n \in \NN$ a sequence $\vec d$ of digits can effectively be found such that $\mu(u, \vec d(x_0)) < 2^{-n}$.
\end{lemma}
\proof
Let $\varepsilon$ be a well-covering number for $(X, D)$ and set
\[
j(n) := \min\set{i \in \NN}{q^{i-1} \cdot M < 2^{-n}}.
\]
Given $u \in Q$ and $n \in \NN$, proceed as follows:
\begin{enumerate}
\item Let $i := 0$.  If $j(n-1) = 0$, output $[]$ (empty sequence). Otherwise, set $v_i := u$, increase $i$ by 1 and go to (2).

\item Assume that $\vec e$ has been computed so far. Use the decidabilty of $(X, D, Q)$ to find some $e \in D$ with $\ball{\mu}{v_i}{\varepsilon} \subseteq e[X]$.
If $j(n-1) = i$, output $\vec{e}e$. Otherwise, use computable density to find some $v_i \in Q$ such that
\[
\mu(v_{i-1}, e(v_i)) < q^{j(n-1) - i} \cdot M/(j(n-1)-1),
\]
increase $i$ by 1 and go to (2).

\end{enumerate}

Now, if $j(n-1) = 0$, we have that $\vec d = []$. Hence,
\[
\mu(u, \vec d(x_0)) = \mu(u, x_0) \le M \le q^{j(n-1)-1} \cdot M < 2^{-n}.
\]
Otherwise, we have found $d_1, \ldots, d_{j(n-1)} \in D$ and $v_0, \ldots, v_{j(n-1)-1} \in Q$ with
\[
\mu(v_k, d_{k+1}(v_{k+1})) < q^{j(n-1)-k-1} \cdot M/(j(n-1)-1),
\]
for $0 \le k < j(n-1)-1$. Then $\vec d = [d_1, \ldots, d_{j(n-1)}]$ and: 
\begin{align*}
\mu(u, \vec d(x_0))
&\le \sum_{k=0}^{j(n-1)-2} \mu(d_1\cdots d_k(v_{k}), d_1 \cdots d_{k+1}(v_{k+1}))\\
 &\hspace{1.8cm}+ \mu(d_1 \cdots d_{j(n-1)-1}(v_{j(n-1)-1}), d_1 \cdots d_{j(n-1)}(x_0))\\
&\le \sum_{k=0}^{j(n-1)-2} q^k \cdot q^{j(n-1)-k-1} \cdot M/(j(n-1)-1) + q^{(j(n-1)-1)} \cdot M\\
& = (j(n-1)-1) \cdot q^{j(n-1)-1} \cdot M/(j(n-1)-1) + q^{(j(n-1)-1)} \cdot M\\ 
&= 2 \cdot q^{j(n-1)-1} \cdot M < 2^{-n} \rlap{\hspace{16.4em} \qedsymbol}
\end{align*}

\begin{lemma}
\label{lem-QD-Q}
Let $(X, D, Q)$ be computable. Then there is a procedure $h$ which given $\vec d \in D^{< \omega}$ and $n \in \NN$ produces a basic element $v \in Q$ so that $\mu(\vec d(x_0), v) < 2^{-n}$.
\end{lemma}
\begin{proof}
Since $x_0 \in \cauchy^\text{eff}$, there is a procedure $f$ which on input $n \in \NN$ computes a basic element $u \in Q$ with $\mu(x_0, u) < 2^{-n}$. Now, define $h$ by recursion on $\vec d$:

On input $n$, if $\vec d = []$, output the result of procedure $f$ on input $n$. Otherwise, assume that $\vec d = e\vec e$ and that the result of $h$ on input $\vec e$ and $n+1$ is $u \in Q$. Then output the result of applying $g_e$ to input $u$ and $n+1$. 
\end{proof}

Summing up we obtain the following result.

\begin{proposition}\label{prop-baseeq}
Let $(X, D, Q)$ be a well-covering, decidable and constructively dense digit space. Then the topological bases $Q$ and $Q_D$ are effectively equivalent.
\end{proposition}

\begin{definition}
\label{def-computable-element}
Let $(X,D)$ be a digit space. An element $x \in X$ is \emph{computable} if there is a computable infinite sequence $\alpha \in D^\omega$ with $\val{\alpha} = x$. Denote the set of all computable elements of $X$ by $X_c$.
\end{definition}

Let $x \in X$ be computable and let this be witnessed by $\alpha \in D^\omega$. With Lemma~\ref{lem-gensur} and (\ref{eq-mudelta}) we obtain that $\mu(x, \alpha^{<n}(x_0)) \le q^n \cdot M$. Set 
\[
\bar{\jmath}(n) := \min\set{i \in \NN}{q^i \cdot M < 2^{-n}},
\]
and assume that $(X, D, Q)$ is computable. By Lemma~\ref{lem-QD-Q}, for any given $n$, we can compute a basic element $v \in Q$ with $\mu(\alpha^{< \bar{\jmath}(n+1)}, v) < 2^{-n-1}$. It follows that $\mu(x, v) < 2^{-n}$. This shows that $X_c \subseteq \cauchy^\text{eff}$. The converse implication will be a consequence of Theorem~\ref{thm-a-sub-c} derived later in a constructive fashion. To this end a further condition is needed.

\begin{definition}
\label{def-continvdeff}
A computable digit space $(X,D,Q)$ \emph{has approximable choice} if for every
$d\in D$ there is an effective procedure 
$\lambda(\theta,u).v^{\theta}_u : \QQ_+\times d[X]\cap Q \to Q$
such that for all $\theta\in\QQ_+$:
\begin{enumerate}

\item\label{def-continvdeff-1} For all $u \in d[X] \cap Q$ and all $\bar{\theta} \in \QQ_+$, $\mu(v^{\theta}_u, v^{\bar{\theta}}_u) < \max\{ \theta, \bar{\theta} \}$.

\item\label{def-continvdeff-2} One can compute $\theta'\in\QQ_+$ such that for all $u,u'\in d[X] \cap Q$,
if $\mu(u, u') < \theta'$ then 
$\mu(v^{\theta}_u,v^{\theta}_{u'}) < \theta$.

\item\label{def-continvdeff-3} For all $u\in d[X]\cap Q$ there is some $z \in d^{-1}[u]$ with 
$\mu(z, v^{\theta}_u) < \theta$.
\end{enumerate} 
\end{definition} 

Obviously,  every computable digit space with approximable choice is constructively dense.

\begin{proposition}
\label{prop-eqcomp-tte}
Let $(X, D, Q)$ be a well-covering and decidable computable digit space with approximable choice. Then $X_c = \cauchy^\text{eff}$.
\end{proposition}

For $\fun{\Phi}{X}{X}$, a map $\fun{\Phi'}{\range(\Phi)}{X}$ is a \emph{right inverse} of $\Phi$ if $\Phi \circ \Phi'$ is the identity on $X$.

\begin {proposition}\label{prop-choice-inv}
A computable digit space $(X, D, Q)$ has approximable choice if, and only if, every $d \in D$ has a computable right inverse.
\end{proposition}
\begin{proof}
Assume that $(X, D, Q)$ has approximable choice and let $d\in D$ and $x \in \range(d)$. Because of density there is some $u_m \in Q \cap \int(d[X])$ with $\mu(x,u_m) < 2^{-m}$, for all $m \in \NN$.

Use approximable choice to pick the function 
$\lambda(\theta,u).v^{\theta}_u$.  
For $\theta_n := 2^{-n-4}$, pick $\theta' \in \QQ_+$ according to approximable choice, part (\ref{def-continvdeff-2}).
Let $N_n \ge 0$ such that $\mu(x, u_m) < \theta'/3$, for $m \ge N_n$. 
Without restriction let $N_n$ be such that $N_n \ge N_i$, for all $i < n$. Set $v_n := v^{\theta_n}_{u_{N_n}}$.
By approximable choice, part (\ref{def-continvdeff-3}), there is some $z_n \in d^{-1}[u_{N_n}]$ with
$\mu(z_n, v_n) < \theta_n$.
Because of the assumption on $N_n$, we have that $\mu(u_{N_m}, u_{N_n}) < \theta'$, 
for $m \ge n$. Hence, $\mu(v_m, v_n) < \theta_n$. It follows that 
$\mu(z_m, z_n) < 3\theta_n < 2^{-n}$. Thus, $(z_n)_{n \in \NN}$ is a fast Cauchy sequence. 
Since $(X,\mu)$ is complete, it converges to some $y_{(u_m)} \in X$. 
As $d$ is continuous, we obtain that 
\[
d(y_{(u_m)}) = \lim_{n \to \infty} d(z_n) = \lim_{n \to \infty} u_n = x.
\]

Now, let $x' \in \range(d)$ with $\mu(x, x') < \theta'/3$ as well as $u'_m \in Q \cap \int(d[X])$ with $\mu(x', u'_m) < 2^{-m}$, for $m \in \NN$. Moreover, $N'_n \ge 0$ such that $\mu(x', u'_m) < \theta'/3$, for $m \ge N'_n$. Without restriction assume that $N'_n \ge N'_i$, for all $i < n$. Finally, let $v'_n := v^{\theta_n}_{u'_{N'_n}}$ and $z'_n \in d^{-1}[u_{N'_n}]$ with $\mu(z'_n, v'_n) < \theta_n$. Then $\mu(u_{N_n}, u'_{N'_n}) < \theta'$ and hence $\mu(v_n, v'_n) < \theta_n$. It follows that $\mu(z_n, z'_n) < 3\theta_n$ and thus $\mu(y_{(u_m)}, y_{(u'_m)}) < 9\theta_n < 2^{-n}$.

For $x=x'$, we obtain that $y_{(u_m)} = y_{(u'_m)}$, ie $y$ does not depend on the choice of the approximating sequence $(u_m)_{m \in \NN}$. Define $d'(x) := y$. By what we have just shown, $d'$ is uniformly continuous with computable modulus of continuity. If $x \in Q \cap \range(d)$, choose $u_m := x$. Then $\mu(d'(x), v_n) < 2\theta_n < 2^{-n}$. Thus, $d'$ is computable.

Conversely, let $d'$ be a right inverse of $d$. For $\theta \in \QQ_+$ let 
\[
m(\theta) := \min\set{m\in \NN}{2^{-m} \le \theta}.
\]
Since $d'$ is computable, we can compute for any given $u \in Q \cap \range(d)$ and $n \in \NN$ a basic element $v \in Q$ so that $\mu(d'(u), v_n) < 2^{-n}$. Set $v^{\theta}_u := v_{m(\theta)+2}$. It remains to verify the conditions in Definition~\ref{def-continvdeff} of having approximable choice:

(\ref{def-continvdeff-1}) Let $\theta, \bar{\theta} \in \QQ_+$. Without restriction let $\theta \ge \bar{\theta}$. Then $\mu(v^{\theta}_u, v^{\bar{\theta}}_u) \le \mu(v^{\theta}_u, d'(u)) + \mu(d'(u), v^{\bar{\theta}}_u) < 2^{-m(\theta)-1} < \theta$.

(\ref{def-continvdeff-2}) As $d'$ has a computable modulus of continuity, for given $\theta \in \QQ_+$ we can compute a $\theta' \in \QQ_+$ such that for $u, u' \in Q \cap \range(d)$, if $\mu(u, u') < \theta'$ then $\mu(d'(u), d'(u')) < \theta/2$, from which it follows that
$\mu(v^{\theta}_u, v^{\theta}_{u'}) < \theta$.

(\ref{def-continvdeff-3}) is obvious: choose $z := d'(u)$.
\end{proof}

In Type-Two Theory of Effectivity an element $x \in X$ is defined to be computable, if it is contained in $A^\text{eff}$. So, it follows that both computability notions coincide. In the present approach elements of $X$ are represented by infinite streams of digits. In Type-Two Theory of Effectivity, similarly, an element $x$ is represented by an infinite sequence $(u_i)_{i \in \NN}$ of basic elements with $\mu(x, u_n) < 2^{-n}$. The resulting representation is called the \emph{Cauchy representation} $\rho_C$. As follows from the proofs leading to the preceding proposition, one can computably pass from an infinite stream $\alpha$ of digits to an infinite sequence of basic elements $(u_i)_{i \in \NN}$ so that $\mu(\val{\alpha}, u_n) < 2^{-n}$, and vice versa. This means that there are computable translations between both representations as summarized by the next result.

\begin{theorem}
\label{thm-stream=cauchy}
Let $(X, D, Q)$ be a well-covering and decidable computable digit space with 
approximable choice. Then there are computable operators 
$\fun{F}{D^\omega}{Q^\omega}$ and $\pfun{G}{Q^\omega}{D^\omega}$ 
such that for $\alpha \in D^\omega$ and $w \in \dom(G)$, 
\[
\rho_C(F(\alpha)) = \val{\alpha} \quad\text{and}\quad \val{G(q)} = \rho_C(w).
\]
\end{theorem}

Let us next start an analogous investigation for compact sets. As has already been mentioned, $(\compact(X), \hdm)$ is a complete compact metric space. It has a canonical dense subset, the set $\QQQ := \powfin(Q)$, the elements of which we will call \emph{basic sets}. 
If $(X, \mu, Q)$ is computable the same holds for $(\compact(X), \hdm, \QQQ)$ (see Brattka~\cite{Brattka99}). 

So, every nonempty compact set is the limit of a fast Cauchy sequence of basic sets with respect to the Hausdorff metric. Brattka and Presser~\cite{BrattkaPresser03} call the resulting representation $\hdr$ of $\compact(X)$ the \emph{Hausdorff representation}.

\begin{proposition}\label{prop-treetohausdorff}
Let $(X, D, Q)$ be a computable digit space. Then there is a 
computable operator 
$\fun{H}{\TTT_D}{\QQQ^\omega}$ such that for every 
$T \in \TTT_D$, $H(T) \in \dom(\hdr)$ with 
\[
\hdr(H(T)) = \val{T}.
\]
\end{proposition}
\begin{proof}
Let $T \in \TTT_D$, $\alpha \in T$ and $n \in \NN$. Moreover, let the computable map $h \colon D^{<\omega} \times \NN \rightarrow Q$ be as in Lemma~\ref{lem-QD-Q}. Then we have that 
\[
\mu(\alpha^{<\bar{\jmath}(n)}(x_0), h(\alpha^{< \bar{\jmath}(n)}, n)) < 2^{-n} \quad\text{and}\quad
\mu(\alpha^{<\bar{\jmath}(n)}(x_0), \alpha^{<\bar{\jmath}(m)}(x_0)) < 2^{-n},
\]
for $m \ge n$. Thus,
\[
\mu(h(\alpha^{< \bar{\jmath}(n+2)}, n+2), h(\alpha^{< \bar{\jmath}(m+2)}, m+2)) < 2^{-n}.
\]
Now, set $H(T) := (C_n)_{n\in\omega}$ where
\[
C_n := \set{h(\vec d, n+2)}{\vec d \in T \cap D^{\bar{\jmath}(n+2)}}.
\]
Then we obtain that $\hdm(C_n, C_m) < 2^{-n}$. In order to see that also $\hdm(\val{T}, C_n) < 2^{-n}$, let $x \in \val{T}$. Hence, $x = \val{\alpha}$, for some $\alpha \in T$. It follows that $\mu(x, \alpha^{< \bar{\jmath}(n)}(x_0)) < 2^{-n}$ and therefore $\mu(x, h(\alpha^{< \bar{\jmath}(n+2)}, n+2)) < 2^{-n}$. Thus, $\inf\set{\mu(x, u)}{u \in C_n} < 2^{-n}$. 

On the other hand, if $u \in C_n$, then $u = h(\gamma^{< \bar{\jmath}(n+2)}, n+2)$, for some $\gamma \in T$. Since $\mu(\val{\gamma}, h(\gamma^{< \bar{\jmath}(n+2)}, n+2)) < 2^{-n}$, we obtain that $\hdm(\val{T}, C_n) < 2^{-n}$. Consequently, $\lim_{n \to \infty} C_n = \val{T}$.
\end{proof}

As a consequence of Theorem~\ref{thm-ak-sub-ck}, also the converse holds: given an infinite word in $\Sigma^\omega$ representing a nonempty compact set $K$, one can compute a digital tree with value $K$.

\begin{theorem}\label{thm-hausdorfftotree}
Let $(X, D, Q)$ be a well-covering and decidable computable digit space 
with approximable choice. Then there are computable operators 
$\fun{H}{\TTT_D}{\QQQ^\omega}$ and 
$\pfun{J}{\QQQ^\omega}{\TTT_D}$ 
such that for $T \in \TTT_D$ and $w \in \dom(J)$,
\[
\hdr(H(T)) = \val{T}\quad\text{and}\quad \val{J(w)} = \hdr(w).
\]
\end{theorem}

It follows that we can effectively translate representations of compact 
sets in Type-Two Theory of Effectivity into representations in our setting, and vice versa.

\begin{definition}
\label{def-computable-compact}
Let $(X,D)$ be a digit space. A set $K\in\compact(X)$ is \emph{computable}
if it is the value of computable digital tree. Denote the set of computable compact sets by $\compact(X)_c$.
\end{definition}

In Type-Two Theory of Effectivity a compact set is \emph{computable} if it is contained in $\cauchy^\text{eff}_K$, where:
\begin{gather*}
\begin{split}
&\cauchy^\text{eff}_K := \{\, K \in \compact(X) \mid \text{there is a procedure that given $n \in \NN$}\\
&\hspace{3cm} \text{computes a basic set $U \in \QQQ$ with $\hdm(K, U) < 2^{-n}$} \,\}
\end{split}
\end{gather*}

With the preceding theorem we obtain that both notions coincide.

\begin{corollary}\label{cor-tte-tree}
Let $(X, D, Q)$ be a well-covering and decidable computable digit space with approximable choice. Then $\compact(X)_c = \cauchy^\text{eff}_K$.
\end{corollary}

\section{Extracting digital trees from coinductive proofs}
\label{sec-coind}
In this section we recast the theory of digit spaces and their hyper-spaces
in a constructive setting with the aim to extract programs that provide
effective representations of certain objects or transformations 
between different representations.
One of the main results will be effective transformations between
the Cauchy--Hausdorff representation and the digital tree representation
of the hyper-space showing that the two representations are effectively
equivalent.
The method of program extraction will be based on a version of realizability,
and the main constructive definition and proof principle will be coinduction.
The advantage of the constructive approach lies in the fact that proofs can be
carried out in a representation-free way. Constructive logic and the Soundness
Theorem guarantee automatically that proofs are witnessed by effective and
provably correct transformations on the level of representations.

Regarding the theory of realizability and its applications to constructive
analysis we refer the reader to Schwichtenberg and Wainer~\cite{SchwichtenbergWainer12},
Berger and Seisenberger~\cite{SeisenBerger10} and Berger~\cite{Berger11}. Here, we only recall the main
facts.
We work in many-sorted first-order logic extended by the formation
of inductive and coinductive predicates.
Realizability assigns to each formula $A$ a unary predicate $\prea{A}$
to be thought of as the set of realizers of $A$. Instead of $\prea{A}(a)$
one often writes $\rea{a}{A}$ (``$a$ realizes $A$'').
The realizers $a$ can be typed or untyped, but for the understanding of the 
following, details about the nature of realizers are irrelevant.
It suffices to think of them as being (idealized, but executable) functional
programs or (Oracle-)Turing machines.
The crucial clauses of realizability for the propositional connectives are:
\begin{align*}
&\rea{c}{(A \lor B)} := \exists a (c=(0,a) \land \rea{a}{A}) \lor 
             \exists b (c=(1,b) \land \rea{b}{B})\\
&\rea{f}{(A\to B)} := \forall a\,(\rea{a}{A}\to \rea{f(a)}{B})\\
&\rea{c}{(A \land B)} := \rea{\pleft{c}}{A} \land \rea{\pright{c}}{B}\\
&\rea{c}{\bot} := \bot
\end{align*}
Hence, an implication is realized by a function and a conjunction is realized
by a pair (accessed by left and right projections, $\pleft{.}, \pright{.}$).

Quantifiers are treated uniformly in our version of realizability:
\begin{align*}
&\rea{a}{(\forall x\,A(x))} := \forall x\,\rea{a}{A(x)}\\
&\rea{a}{(\exists x\,A(x))} := \exists x\,\rea{a}{A(x)}
\end{align*}
This reflects the fact we allow variables $x$ to range over abstract 
mathematical objects without prescribed computational meaning. Therefore, the
usual interpretation of $\rea{a}{(\forall x\,A(x))}$ to mean
$\forall x\,\rea{a(x)}{A(x)}$ doesn't make sense since we would use
the abstract object $x$ as an input to the program $a$.

For atomic formulas $P(\vec t)$, where $P$ is a predicate and $\vec t$ are 
terms, realizability is defined in terms of a chosen predicate $\reap{P}$ 
with on extra argument place, that is,
\begin{align*}
&\rea{a}{P(\vec t)} := \reap{P}(a,\vec t).
\end{align*}
The choice of the predicates $\reap{P}$ allows us to fine tune the computational
content of proofs.

So far, we have covered first-order logic. Now we explain how inductive and 
coinductive definitions are realized. 
An inductively define predicate $P$ is defined as the least fixed point of
a monotone predicate transformer $\Phi(X,\vec x)$, that is the formula
$\forall\vec x(X(\vec x) \to Y(\vec x)) \to 
\forall \vec x(\Phi(X,\vec x) \to \Phi(Y,\vec x)$,
with free predicate variables $X$ and $Y$, must be provable.
Then we have the closure axiom
\[  \forall\vec x(\Phi(P,\vec x) \to P(\vec x))\]
as well as the induction schema
\[   \forall\vec x(\Phi(\mathcal{A},\vec x) \to \mathcal{A}(\vec x))
     \to \forall\vec x(P(\vec x) \to \mathcal{A}(\vec x)) \]
for every predicate $\mathcal{A}$ defined by some formula $A(\vec x)$ 
as $\mathcal{A}(\vec x) \leftrightarrow A(\vec x)$.
Realizability for $P$ is defined by defining $\reap{P}$ inductively by 
the operator 
$\reap{\Phi}(\tilde{X},a,\vec x) := \rea{a}{\Phi(X,\vec x)}$.
This means we have the closure axiom 
\[  \forall a,\vec x(\rea{a}{\Phi(P,\vec x)} \to \reap{P}(a,\vec x))\]
as well as the induction schema:
\[   \forall a,\vec x(\rea{a}{\Phi(\mathcal{A},a,\vec x)} 
     \to \mathcal{A}(a,\vec x))
     \to \forall a,\vec x(\reap{P}(a,\vec x) \to \mathcal{A}(a,\vec x)) \]
Dually, $\Phi$ also gives rise to a coinductively defined predicate $Q$ 
defined as the greatest fixed point of $\Phi$. Hence, we have the coclosure axiom
\[  \forall\vec x(Q(\vec x) \to \Phi(Q,\vec x))\]
and the coinduction schema:
\[   \forall\vec x(\mathcal{A}(\vec x) \to \Phi(\mathcal{A},\vec x))
     \to \forall\vec x(\mathcal{A}(\vec x) \to Q(\vec x)) \]
Realizability for $Q$ is defined by
defining $\reap{Q}$ coinductively by the same operator $\reap{\Phi}$ as above,
hence, the coclosure axiom 
\[  \forall a,\vec x(\reap{Q}(a,\vec x) \to \rea{a}{\Phi(Q,\vec x)})\]
and the coinduction schema:
\[   \forall a,\vec x(\mathcal{A}(a,\vec x))
     \to \rea{a}{\Phi(\mathcal{A},a,\vec x)} 
     \to \forall a,\vec x(\mathcal{A}(a,\vec x) \to \reap{Q}(a,\vec x)) \]
The basis for program extraction from proofs is the Soundness Theorem.
\begin{sdthm}
 From a constructive proof of a formula $A$
from assumptions $B_1,\ldots,B_n$ one can extract a program 
$M(a_1,\ldots, a_n)$ such that $\rea{M(a_1,\ldots,a_n)}{A}$ is provable
from the assumptions $\rea{a_1}{B_1},\ldots,\rea{a_n}{B_n}$.
\end{sdthm}
 
If one wants to apply this theorem to obtain a program realizing the
formula $A$ one must provide terms $K_1,\ldots,K_n$ realizing the assumptions
$B_1,\ldots,B_n$. Then it follows that the term $M(K_1,\ldots,K_n)$ realizes
$A$.

That realizers do actually \emph{compute} witnesses is shown in 
Berger~\cite{Berger10} and Berger and Seisenberger~\cite{SeisenBerger10} by a 
\emph{Computational Adequacy Theorem} that relates the denotational 
definition of realizability with a lazy operational semantics. 

There is an important class of formulas where realizers do not matter:
We call a formula $B$ \emph{non-computational} if
\[\forall a(\rea{a}{B} \leftrightarrow B)\] 
is provable.
Now, if we have proven $A$ from assumptions $B_1,\ldots,B_n$, where 
$B_1,\ldots, B_k$ are non-computable, then we can extract a realizer 
of $A$ that depends only on realizers of $B_{k+1},\ldots,B_n$ and whose
correctness can be proven from the assumptions $B_1,\ldots,B_k$.
We can simplify the definition of realizability for formulas with 
non-computational parts: If $A$ is non-computable, then:
\begin{align*}
&\rea{a}{A} := A\\
&\rea{b}{(A \land B)} := A \land \rea{a}{B}\\
&\rea{b}{(A\to B)} := A\to \rea{b}{B}
\end{align*}
Non-computational formulas can simplify program extraction drastically.
Therefore, it is important to have handy criteria for recognizing 
non-computable formulas.
First of all, realizability, and hence the question which formulas are 
non-computable, depends on how the predicates 
$\reap{P}$, defining realizability of atomic formulas $P(\vec t)$, are
axiomatized. We call a predicate $P$ non-computational if the 
axiom for $\reap{P}$ is:
\[ \forall a,\vec x(\reap{P}(a,\vec x) \leftrightarrow P(\vec x)) \]
Now, clearly, $P(\vec t)$ is non-computable, if $P$ is non-computable. Furthermore, 
$\bot$ is non-computable and it is easy to see that the set of non-computable formulas is 
closed under implication, conjunction and universal and existential 
quantification.
Moreover, if $A$ is a \emph{faithful} formula, that is,
\[ A \leftrightarrow \exists a (\rea{a}{A})\]
then $A\to B$ is non-computable, provided $B$ is non-computable. In particular, the negation 
of a faithful formula is non-computable.
Clearly, every non-computable formula is faithful and it is easy to see that
the set of faithful formulas is closed under conjunction, disjunction and 
existential quantification.

In our formalization and realizability interpretation of the theory of real
numbers and digit spaces, we regard the set $\RR$ of real numbers as well as
the carrier set $X$ of an arbitrary but fixed metric space $(X,\mu)$ as sorts.
All arithmetic constants and functions we wish to talk about as well as the
metric $\mu$ are admitted as constants or function symbols.  We
declare the predicates $=$, $<$ and $\le$ on $\RR$ as non-computational.
Furthermore, we admit all true non-computable statements about real numbers as well as
the axioms of a metric space (which are non-computable formulas) as axioms in our
formalization.

In order to be able to formalize a digit space $(X,D)$ and the hyper-space
$\compact(X)$ we add to every sort $s$ its powersort $\pow{s}$,
equipped with a non-computational element-hood relation $\epsilon$,
as well as a
function space sort $s\to t$ to any two sorts $r$ and $t$,
equipped with an application operation and operations such as composition
and the like. Furthermore, we add for every non-computable formula $A(x)$ the 
comprehension axiom
\[ \exists u\, \forall x\, (x \epsilon u \leftrightarrow A(x)) \]
($A(x)$ may contain other free variables than $x$).
We will use the notation $\set{x}{A(x)}$ for the element $u$ of sort
$\pow{s}$ whose existence is postulated in the comprehension axiom above.
Hence, we can define the empty set $\emptyset := \set{x}{\bot}$,
singletons $\{y\} := \set{x}{x = y}$ and the classical union of two sets
$u\cup v := \set{x}{\neg\neg (x \epsilon u \lor x \epsilon v)}$.

The comprehension axiom above is an example of an non-computable formula which we
wish to accept as true. In general, we may admit any non-computable formula
as axiom which is true in an intended model or provable in some
accepted theory (which may be classical).

Our first example of an inductive definition is the predicate $\powf{P}$ of
finite subsets of a predicate $P$
\begin{align*}
& \powf{P}(\emptyset)\\
& \text{If $\powf{P}(v)$ and $P(x)$, then $\powf{P}(v \cup \{x\})$}
\end{align*}
(Here $v$ is a variable of sort $\pow{s}$ and $x$ is a variable of sort $s$).
More formally, $\powf{P}$ is defined as the least fixed point of the operator:
\[\Phi(X,u) := u = \emptyset \lor \exists v\,\exists x\,(X(v) \land 
                                          u =  v \cup \{x\}) \]
Above, we may view $\emptyset$ and $\cup$ as a new constant and function
symbol, or else eliminate them with the usual technique (as in set theory).

The next example is a coinductive predicate that generalizes a corresponding
definition of a predicate $\coco_0$ on the signed digit space $(\II,\AV)$
introduced in Berger~\cite{Berger11} and Berger and Seisenberger~\cite{SeisenBerger10}.
\begin{definition}
\label{def-coco}
Let $(X,D)$ be a digit space.
We define coinductively $\coco_0$ as the largest subset of $X$ such that 
for all $x \in X$:
\[
x \in \coco_0 \Rightarrow \exists d \in D\, \exists y \in \coco_0\,\, x = d(y)
\]
\end{definition}

\begin{lemma}
\label{lem-coco-space}
$\coco_0 = X$.
\end{lemma} 
\begin{proof}
By definition, $\coco_0 \subseteq X$. For the proof of the converse inclusion 
it suffices to observe that because of Proposition~\ref{prop-valcont}(\ref{valcont1}) and Lemma~\ref{lem-valrec} the defining implication of $\coco_0$ remains
correct if $\coco_0$ is replaced by $X$.
\end{proof}
Hence, classically, the set $\coco_0$ is rather uninteresting, 
but, constructively, it is significant, since 
from a constructive proof that $x\in\coco_0$ one can extract a  
stream of signed digits $\alpha$ such that $x = \val{\alpha}$.
Furthermore, as shown by Berger~\cite{Berger11}, in the case of 
the signed digit $(\II,\AV)$, one can extract
from a constructive proof that $\coco_0$ is closed under, say, 
multiplication, a program for multiplication with respect
to the signed digit representation.

In this paper we investigate whether what was done for the \emph{points} of
$\II$ can be done in a similar way for the \emph{nonempty compact subsets} of 
$\II$, (or, more generally, for the nonempty compact subsets of the 
underlying space $X$ of a digit space $(X,D)$).
\begin{definition}
\label{def-coco-compact}
Given a digit space $(X,D)$, we define, coinductively, the set 
$\coco_{\compact}$ as the largest subset of $\compact(X)$ such that:
\[ A \in \coco_{\compact} \Rightarrow
    \exists E \subseteq D\, \exists (A_d)_{d\in E} \in \compact(X)^E \ 
  (A = \bigcup_{d\in E} d[A_d] \land \forall d\in E\,A_d\in\coco_{\compact})\]
\end{definition}
\begin{lemma}
\label{lem-coco-compact}
$\coco_{\compact}=\compact(X)$.
\end{lemma}
\begin{proof}
By definition, $\coco_{\compact}\subseteq\compact(X)$.
The converse inclusion follows by coinduction, since
the implication in the above statement holds by Lemma~\ref{lem-tree-compact} and Corollary~\ref{cor-tree-valrec}.
\end{proof}

The proof of Lemma~\ref{lem-coco-compact} is classical because the set
$E$ cannot be computed since, in general, one cannot decide whether 
$d[X]\cap A\neq\emptyset$.
The significance of the definition of $\coco_{\compact}$ stems from the 
fact that the realizers of a statement $A\in\coco_{\compact}$ are 
exactly the digital trees representing $A$, as we will show below. 
It follows from the definition of realizability in Berger~\cite{Berger11}
that the type $\tau$ of realizers of a statement $A\in\coco_{\compact}$ 
is defined by the recursive type equation
\[ \tau = \Sigma_{E\in\powplus(D)} \tau^{\card{E}} \] 
where $\powplus(D)$ is the set of nonempty subsets of $D$ and $\card{E}$ is the 
cardinality of $E$. Using 
constructive terminology one would call $\powplus(D)$ the set of 
decidable inhabited subsets of $D$; since $D$ is finite this set is 
finite as well and its cardinality exists constructively.
For example, if $D$ has three elements $i,j,k$, then:
\[ \tau = \{i\}\times\tau +\{j\}\times\tau +\{k\}\times\tau +
\{i,j\}\times\tau^2 +\{i,k\}\times\tau^2 + \{j,k\}\times\tau^2 +
\{i,j,k\}\times\tau^3\]
One sees that $\tau$ is, essentially, the set $\TTT_D$ of digital
trees. Indeed, every digital tree $T$ can be identified with the pair
$(\branches{T},(\subtree{T}{d})_{d\in E})$ where 
$\branches{T} := \{\,d\in D\mid [d]\in T\,\}$ and
$\subtree{T}{d} := \{\,\vec{d}\in D^* \mid d\vec{d}\in T\,\}$. 
Note that $\val{\subtree{T}{d}} = d^{-1}[\val{T}]$ 
(since $\alpha$ is a path in $\subtree{T}{d}$ exactly if $d\alpha$ is a path in $T$,
and $\val{d\alpha} = d(\val{\alpha})$).
What it means for a digital tree $T$ to realize that 
$A\in\coco_{\compact}$, written $\rea{T}{(A\in\coco_{\compact})}$,
is defined coinductively as the largest subset $\rea{\cdot}{\cdot}$ of
$\TTT_D\times\compact(X)$ such that if
$\rea{T}{(A\in \coco_{\compact})}$, then:
\[ 
  \exists (A_d)_{d\in \branches{T}} \in \compact(X)^{\branches{T}} \, 
(A = \bigcup_{d\in \branches{T}} d[A_d] \land 
\forall d\in \branches{T}\,\rea{\subtree{T}{d}}{(A_d\in\coco_{\compact})})
\]

\begin{theorem}
\label{thm-coco-compact-real}
The realizers of a statement $A\in\coco_{\compact}$ are exactly the 
digital trees representing $A$, that is: 
\[\rea{T}{(A\in\coco_{\compact})} \Longleftrightarrow \val{T}=A \]
In particular, from a constructive proof of $A\in\coco_{\compact}$ one 
can extract a digital tree representation of $A$.
\end{theorem}
\begin{proof}
We first show by coinduction that if $\val{T}=A$ then 
$\rea{T}{(A\in\coco_{\compact})}$.
This means we have to show that the implication defining 
the relation $\rea{\cdot}{\cdot}$ holds if that relation
is replaced by the relation $\val{\cdot} = \cdot$.
Hence assume $\val{T} = A$. 
For $d\in \branches{T}$ set 
$A_d := d^{-1}[A]$ which is compact and nonempty (since $d\in \branches{T}$).
Furthermore, $A = \bigcup_{d\in \branches{T}} d[A_d]$ and
$\val{\subtree{T}{d}} = d^{-1}[\val{T}] = d^{-1}[A] = A_d$.
For the converse implication it suffices to show that any realizer
of a statement $A\in\coco_{\compact}$ has a value which is arbitrarily 
close to $A$ in the Hausdorff metric. More precisely we show
\[\forall n\in\NN\,\forall T\in\TTT_X\forall A\in\compact(X)\,(\,
    \rea{T}{(A\in\coco_{\compact})} \Rightarrow \hdm(\val{T},A)\le M\cdot 
q^{n})\]
by induction on $n$, where $M$ is a bound for $X$ and $q<1$ is a common 
contraction factor for the digits in $D$. 
The case $n=0$ is trivial. 
For the step we assume $\rea{T}{(A\in\coco_{\compact})}$ and show 
$\hdm(\val{T},A)\le M\cdot q^{n+1}$.
By the assumption we have 
$(A_d)_{d\in \branches{T}} \in \compact(X)^{\branches{T}}$ such that
$A = \bigcup_{d\in \branches{T}} d[A_d]$ and 
$\rea{\subtree{T}{d}}{(A_d\in\coco_{\compact})}$ for all $d\in\branches{T}$.
By the induction hypothesis we have 
$\hdm(\val{\subtree{T}{d}},A_d)\le M\cdot q^{n}$ for all $d\in\branches{T}$.
Since all digits are contracting by the factor $q$, it follows that
$\hdm(d[\val{\subtree{T}{d}}],d[A_d])\le M\cdot q^{n+1}$ for all 
$d\in\branches{T}$. 
Since $\val{T} = \bigcup_{d\in \branches{T}}d[\val{\subtree{T}{d}}]$ and
$A = \bigcup_{d\in \branches{T}} d[A_d]$, we conclude that
$\hdm(\val{T},A)\le M\cdot q^{n+1}$.
\end{proof}

\section{Extracting the Cantor Set}
\label{sec-cantor}
As an example, we prove that the Cantor set  
lies in $\coco_{\compact}$ and extract a program
that computes a digital tree representation of it.
Of course, in this example, the digit space under discussion is the signed
digit space $(\II,\AV)$ which was introduced in Sect.~\ref{sec-digit-space}.
For convenience, we consider a scaled version of the Cantor set 
that fits better with
the signed digits. Therefore, we define the Cantor set $\cantor$
as the fractal defined by the contractions: 
\[\fminus(x) = \frac{x-2}{3}, \quad \fplus(x) = \frac{x+2}{3}\]
More precisely, $\cantor$ is the unique attractor of the iterated 
function system (IFS) $\{\fminus,\fplus\}$ and can be defined explicitly
as
\[\cantor := \bigcap_{n\in\NN} F^n(\II) \]
where $F(A) := \fminus[A] \cup \fplus[A]$.
Intuitively, this means that we start with the interval $\II$,
remove the open middle third and repeat the process with the 
remaining parts\footnote{Traditionally, one would start with the interval $[0,1]$
and use the maps $x\mapsto x/3$ and $x\mapsto(x+2)/3$.}.
The only facts we will be using about the set $\cantor$ are that 
it is a nonempty subset of $\II$ and a fixed point of $F$.

\begin{theorem}
\label{thm-cantor}
$\cantor\in\coco_{\compact}$.
\end{theorem}
\begin{proof}
By a \emph{positive affine linear map (palm)} we mean a real function of 
the form $h(x) = ux + v$ where $u,v\in\RR$ with $u > 0$. 
The set of all palms is a subgroup of the permutation group of 
$\RR$, ie
palms are closed under composition and they are bijective with their 
inverses again being palms. 
The maps $\fminus$ and $\fplus$ as well as the signed digit maps 
$\av{i}\in\AV$ are examples of palms that map $\II$ into itself.
Note that for a palm $h(x)=ux+v$ we have
$h[\II]\subseteq\II$ exactly if $u+|v|\le 1$.
A \emph{rational palm} is a palm with rational coefficients $u,v$. 
The rational palms form a subgroup of
all palms. The examples above are rational palms and in the following we will
be working exclusively with rational palms.

In order to show that $\cantor\in\coco_{\compact}$ we show more generally,
by coinduction that the set
\[
\cantor' := \{\,h[\cantor]\mid h\hbox{ a rational palm with }
                                h[\II]\subseteq\II\,\} 
\]
is a subset of $\coco_{\compact}$. 
By the definition of $\coco_{\compact}$ we have to show:
\[ A \in \cantor' \Rightarrow A \neq\emptyset \land 
    \exists E \subseteq \AV\, \exists (A_d)_{d\in E} \in (\cantor')^E \ 
      A = \bigcup_{d\in E} d[A_d] \]
Since all elements of $\cantor'$ are nonempty subsets of $\II$ 
(since $\cantor$ is nonempty), this amounts to showing that
for every rational palm $h$ with $h[\II]\subseteq\II$ we can find a set 
$E\subseteq\{-1,0,1\}$ and rational palms $h_i$ ($i\in E$) with 
$h_i[\II]\subseteq\II$ such that: 
\[h[\cantor] = \bigcup_{i\in E} \av{i}[h_i[\cantor]] \] 
We first consider the easy case that $h[\II]\subseteq\av{i}[\II]$ for some
$i\in\{-1,0,1\}$. In that case we can take $E := \{i\}$ and 
$h_i := \av{i}^{-1}\circ h$, since 
$h[\cantor] = \av{i}[(\av{i}^{-1}\circ h)[\cantor]]$
and $(\av{i}^{-1}\circ h)[\II] = \av{i}^{-1}[h[\II]] \subseteq 
\av{i}^{-1}[\av{i}[\II]] = \II$.

If we are not in that easy situation, we choose $i,j\in\{-1,0,1\}$ such that
$(h\circ \fminus)[\II] \subseteq \av{i}[\II]$ and 
$(h\circ \fplus)[\II] \subseteq \av{j}[\II]$ (that such $i,j$ do exist will be 
shown later). We set $E := \{i,j\}$ and 
\[ h_{i} := \av{i}^{-1}\circ h\circ \fminus,\qquad
   h_{j} := \av{j}^{-1}\circ h\circ \fplus.\]
Using the fact that $\cantor$ is a fixed point of $F$, ie
\[\cantor = \fminus[\cantor] \cup \fplus[\cantor],\]
we obtain:
\[ h[\cantor] = (h\circ\fminus)[\cantor] \cup (h\circ\fplus)[\cantor]
 = \av{i}[h_i[\cantor]] \cup \av{j}[h_j[\cantor]]
\]
Furthermore, by the choice of $i,j$, we have, as required,
\[ h_i[\II] = \av{i}^{-1}[(h\circ\fminus)[\II]] \subseteq 
\av{i}^{-1}[\av{i}[\II]] = \II,\]
and similarly, $h_j[\II] \subseteq \II$.

It remains to be shown that $i,j$ above can always be found. 
First, note 
\[\av{-1}[\II] = [-1,0],\quad \av{0}[\II] = [-1/2,1/2], \quad
\av{1}[\II] = [0,1].\]
Assume $h(x) = ux+v$ and let $a:= h(-1) = v-u \ge -1$ and 
$b := h(1) = v+u \le 1$. Then $h[\II]=[a,b]\subseteq \II$. 
Since we are not in the easy case, we may assume $a < 0 < b$. 
To determine $i$, note that
\[(h\circ\fminus)[\II] = [h(\fminus(-1)),h(\fminus(1))] = [a,v-u/3]. \]
Consider the case $a\le -1/2$. Then we have  $v \le 1/4$, since 
if $v > 1/4$ we would have $u = v - a > 3/4$ and therefore $b = v + u > 1$.
It follows that $v - u/3 = (2v + a)/3  \le 0$. This means that we can take
$i = -1$. 
In the case $a \ge -1/2$ we can take $i = 0$ since $v - u/3 \le v \le 1/2$
(if $v > 1/2$ we would have $u = v - a \ge v > 1/2$ and therefore $b > 1$).
The calculation of $j$ is symmetric: if $b\ge 1/2$, we can take $j=1$, 
otherwise $j = 0$.
\end{proof}

\section{Equivalence with the Cauchy representation}
\label{sec-equiv}

Let $(X,D,Q)$ be a computable digit space.
We define the predicate $\cauchy\subseteq X$ by:
\[ \cauchy(x) \equiv 
   \forall n\in\NN\,\exists u\in Q\,\mu(x,u) < 2^{-n} \]
A realizer of $\cauchy(x)$ is a fast Cauchy sequence in $Q$ converging to $x$.

\begin{theorem}
\label{thm-c-sub-a}
$\coco_0 \subseteq \cauchy$.
\end{theorem}
\begin{proof}
Because of Lemma~\ref{lem-QD-Q} it suffices to show that
\[
\forall n\in\NN\,\forall x\in X\,(\coco_0(x) \Rightarrow 
   \exists u\in Q_D\,\mu(x,u) \le M \cdot q^n \,),
   \]
which will be done by induction on $n$.
If $n=0$, let $u$ be any element in $Q_D$.
For $n+1$, assume $x\in\coco_0$. Then there are 
$d\in D$ and $y\in\coco_0$ such that $x=d(y)$. By induction hypothesis,
there exists $v\in Q_D$ such that $\mu(y,v)\le M\cdot q^{n}$. 
Set $u := d(b)\in Q_D$. Then, $\mu(x,u) \le q\cdot\mu(y,v) \le M\cdot q^{n+1}$. 
\end{proof}

\begin{theorem}
\label{thm-a-sub-c}
Let $(X,D,Q)$ be a well-covering and decidable computable digit space with approximable choice.
Then $\cauchy \subseteq \coco_0$.
\end{theorem}
\begin{proof}
We prove the statement by coinduction. Hence assume $\cauchy(x)$.
We have to find $d\in D$ and $y\in X$ such that $x=d(y)$ and $\cauchy(y)$.
Let $\varepsilon \in \QQ_+$ be a well-covering number.
Using $\cauchy(x)$,  
pick $\hat{u}\in Q$ such that $\mu(x,\hat{u}) < \varepsilon/2$.
Pick $d\in D$ such that $\ball{\mu}{\hat{u}}{\varepsilon}\subseteq d[X]$. 
Then $x \in \ball{\mu}{\hat{u}}{\varepsilon}$. 

By Proposition~\ref{prop-choice-inv}, $d$ has a computable right inverse $d'$. 
Set $y := d'(x)$. Since $d'$ has a computable modulus of continuity, we can, 
given $n \in \NN$, compute a number $k(n)$ so that for $x', x'' \in d[X]$, 
if $\mu(x', x'') < 2^{-k(n)}$ then $\mu(d'(x'), d'(x'')) < 2^{-n-1}$. 
Using the assumption $\cauchy(x)$ again, we find $u\in Q$ such that
$\mu(x,u)< 2^{-k(n)}$.
It follows that $\mu(d'(x), d'(u))) < 2^{-n-1}$. By the computability 
of $d'$ we can moreover compute a basic element $v \in Q$ with 
$\mu(d'(u), v) < 2^{-n-1}$. 
Hence, $\mu(y, v) < 2^{-n}$, which shows that $A(y)$.
\end{proof}

Now we do for the hyper-space $\compact(X)$ what we did for $X$ above.
We define the predicate $\cauchy_\compact\subseteq \compact(X)$ by:
\[ \cauchy_\compact(A) \equiv 
   \forall n\in\NN\,\exists U\in \powfin(Q)\ \hdm(A,U) < 2^{-n} \]
A realizer of $\cauchy_\compact(A)$ is a fast Cauchy sequence of nonempty 
finite subsets of $Q$ converging to $A$.

\begin{theorem}
\label{thm-ck-sub-ak}
$\coco_\compact \subseteq \cauchy_\compact$.
\end{theorem}
\begin{proof}
We show 
\[
\forall n\in\NN\,\forall A\in \compact(X)\,(\coco_\compact(A) \Rightarrow 
  \exists U\in \powfin(Q)\,\hdm(A,U) \le M \cdot q^n\,),
  \]
by induction on $n$.
If $n=0$, we can take $U := \{u\}$, for any $u \in Q$.
For $n+1$, assume $A\in\coco_\compact$. Then there are 
$E \subseteq D$ and a family $(A_d)_{d\in E} \in \compact(X)^E$
such that $A = \bigcup_{d\in E} d[A_d]$ and $A_d\in\coco_{\compact}$ for
all $d\in E$.
By induction hypothesis,
there exist $V_d\in \powfin(Q)$ such that $\hdm(A_d,V_d)\le M\cdot q^{n}$,
for all $d\in E$. Set $U := \bigcup\{\,d[V_d] \mid d\in E\,\}$. 
Then it follows with Lemma~\ref{lem-hausm} that $\hdm(A,U) \le \max\{\,q\cdot\hdm(A_d,V_d)\mid d\in E\,\} 
\le M\cdot q^{n+1}$. 
\end{proof}

\begin{theorem}
\label{thm-ak-sub-ck}
Let $(X,D,Q)$ be a well-covering and decidable computable digit space with approximable choice. Then $\cauchy_{\compact} \subseteq \coco_{\compact}$.
\end{theorem}
The proof will be based on a sequence of intermediate results.

We say that $(X,D,Q)$ has \emph{property (P)} if the 
following holds: For every $K \in \compact(X)$ and every sequence 
$(C_n)_{n\in\NN}$ of basic sets
with $\hdm(C_n,K) < 2^{-n}$ for all $n$, one can compute from $(C_n)$
\begin{itemize}
\item[-] a decision procedure for a set $E$ of digits in $D$,
\item[-] for every $d \in E$ a sequence of basic sets $(C^d_n)_{n\in\NN}$
\end{itemize}
such that there exist $K_d \in \compact(X)$ for each $d \in E$, with
$K = \bigcup \{\,d[K_d] \mid d \in E\,\}$ and $\hdm(C^d_n,K_d) < 2^{-n}$ for all 
$n$.

An equivalent way of stating the property (P) is to state constructively: 
if $\cauchy_\compact(A)$ holds, then there exists a decidable set 
$E\subseteq D$ and $A_d \in \compact(X)$ with $\cauchy_\compact(A_d)$ 
for each $d \in E$. 

\begin{lemma}
\label{lem-p}
If $(X,D,Q)$ has property (P), then 
$\cauchy_\compact \subseteq\coco_\compact$.

In terms of realizers this means: for every set $K \in \compact(X)$ 
and every sequence $(C_n)$ of basic sets with $\hdm(C_n,K) < 2^{-n}$ for all 
$n$, one can compute from $(C_n)$ a digital tree $T$ such that $\val{T} = K$.
\end{lemma}
\begin{proof}
Immediate, by coinduction.

Alternatively, one can directly define a decision procedure for a tree defining
a set $K\in\compact(X)$ from a Cauchy-sequence of basic sets converging to $K$.
To this end, assume that $\hdm(K,C_n)<2^{-n}$ for all $n\in\NN$ where the 
$C_n$ are basic sets. We define a function 
\[t : D^{\le\omega} \to (\compact(X)\times\basic(X)^\omega)\cup\{\ast\}\]
where $\basic(X)$ is the set of basic sets and $\ast$ is a new symbol
meaning intuitively ``not in the tree''. The definition of $t(\vec d)$ is
by recursion on $\vec d$ and will be such that whenever 
$t(\vec d) = (K',(C'_n)_{n\in\NN})$, then $\hdm(K',C'_n)<2^{-n}$ for all 
$n\in\NN$.
\begin{itemize}
\item[-] $t([]) := (K,(C_n)_{n\in\NN})$
\item[-] If $t(\vec d) =\ast$, then $t(\vec dd) := \ast$
\item[-] If $t(\vec d) = (K',(C'_n)_{n\in\NN})$, then we use 
property (P), with $K'$ and $(C'_n)_{n\in\NN}$, to compute $E\subseteq D$
and for every $d\in E$ a sequence of basic sets $(C'^d_n)_{n\in\NN}$ such that 
there exist $K'_d\in\compact(X)$ for $d\in E$ with 
$K'=\bigcup_{d\in E}d[K'_d]$ and $\hdm(K'_d,C'^d_n)<2^{-n}$ for all $n\in\NN$.
\begin{itemize}
\item[-] If $d\not\in E$, then $t(\vec dd) := \ast$
\item[-] If $d\in E$, then $t(\vec dd) := (K'_d,(C'^d_n)_{n\in\NN})$
\end{itemize}
\end{itemize}
Finally we define $T := \{\,\vec d \in D^\omega \mid t(\vec d) \neq \ast\,\}$.
Clearly, $T$ is a digital tree that can be computed from the sequence
$(C_n)_{n\in\NN}$ (more precisely, a decision procedure for $T$ can be
computed from $(C_n)_{n\in\NN}$). We show that $\val{T} = K$.

Let $\alpha$ be a path in $T$, that is, 
$t(\alpha^{< m})=(K_m,(C^m_n)_{n\in\NN})$ for all $m\in\NN$.
In order to show $\val{\alpha}\in K$ it suffices to show
that $\alpha^{< m}[X]\cap K\neq\emptyset$ for all $m\in\NN$.
We show that $\alpha^{< m}[K_m]\subseteq K$, by induction on $m$. 
$\alpha^{< 0}[K_0] = K_0 = K$.
$\alpha^{< m+1}[K_{m+1}] = \alpha^{< m}[\alpha_m[K_{m+1}]] \subseteq
\alpha^{< m}[K_{m}]] \subseteq K$, by induction hypothesis, and since,
by construction, $\alpha_m[K_{m+1}] \subseteq K_m$.

Conversely, let $x\in K$. We define recursively $\alpha\in D^\omega$,
such that for all $m\in\NN$, $t(\alpha^{<m})\neq\ast$, hence
$t(\alpha^{<m})$ is of the form $(K_m,\ldots)$ and $x\in\alpha^{<m}[K_m]$.
It follows then that $\alpha\in T$ and $x=\val{\alpha}$.
For $m=0$ there is nothing to define since 
$t(\alpha^{<0}) = t([]) = (K,\ldots)$ and $x\in K$, by assumption.
Now suppose $\alpha^{<m}$ has been defined such that 
$t(\alpha^{<m}) = (K_m,\ldots)$ and $x\in\alpha^{<m}[K_m]$.
Let $y\in K_m$ such that $x = \alpha^{<m}(y)$. By the definition of $t$,
we have a set $E\subseteq D$ such that $\alpha^{<m}d\in T$ exactly if $d\in E$, 
as well as $K_m = \bigcup_{d\in E} d[K^d]$ for some 
$K^d\in\compact(X)$ such that for $d\in E$, $t(\alpha^{<n}d)=(K^d,\ldots)$. 
Hence $y\in d[K^d]$ for some $d\in E$. 
Set $\alpha(m) := d$ and $K_{m+1} := K^d$. It follows that 
$t(\alpha^{<m+1})$ is of the form $(K_{m+1},\ldots)$ and 
$x\in \alpha^{<m+1}[K_{m+1}]$.
\end{proof}
\begin{lemma}
\label{lem-xink}
Let $(X,\mu)$ be a compact metric space.
Let $K\in\compact(X)$ and $C_n\in\compact(X)$
such that $\hdm(K,C_n)\le 2^{-n}$ for all $n\in\NN$.
\begin{itemize}
\item[(a)] For $x\in X$ the following are equivalent:
\begin{itemize}
\item[(i)] $x\in K$
\item[(ii)] $\forall n\in \NN\,\exists y\in C_n\,\mu(x,y)< 2^{-n}$
\item[(iii)] $\forall n\in \NN\,\exists k\ge n\,\exists y\in C_k\,
               \mu(x,y)< 2^{-n}$
\end{itemize}
\item[(b)] Let $K'\in\compact(X)$ and $C'_n\in\compact(X)$
such that $\hdm(K',C'_n)\le 2^{-n}$ for all $n\in\NN$.
If for every $n\in\NN$ there exists $k\ge n$ with $C'_n\subseteq C_k$,
then $K' \subseteq K$.
\end{itemize}
\end{lemma}
\begin{proof}
(a) The implications from (i) to (ii) and from (ii) to (iii) are trivial.
Assume (iii) holds. For every $n\in\NN$ let $k_n\ge n$ and $y_n\in C_{k_n}$
such that $\mu(x,y_n)<2^{-n}$ (using assumption (iii)) and also let
$x_n\in K$ such that $\mu(y_n,x_n) < 2^{-k_n}$ (using the hypothesis that
$\hdm(K,C_{k_n})\le 2^{-k_n}$).
It follows $\mu(x,x_n) \le 2^{-n} + 2^{-k_n} \le 2^{1-n}$.
Hence $x$ is a limit of points in $K$ and therefore in K, since $K$ is closed.

(b) Let $x \in K'$. From the assumptions it easily follows that condition (iii)
of part (a) holds. Therefore $x\in K$.
\end{proof}

\begin{lemma}
\label{lem-split}
Let $(X,\mu)$ be a compact metric space with dense subset $Q$.  Let $D$ be a
finite index set and $X_d$ a nonempty closed subset of $X$ for each $d\in
D$. Assume that the $X_d$ well-cover $X$, ie there exists a rational 
$\varepsilon>0$ such that for each $x\in X$ there exists $d\in D$ with
$\ball{\mu}{x}{\varepsilon}\subseteq X_d$.

Let $K\in\compact(X)$ and $(C_n)_{n\in\NN}$ a family of basic sets such that 
$\hdm(K,C_n)\le 2^{-n}$ for all $n\in\NN$.

Then there exists $E\subseteq D$ and for each $d\in E$ a set 
$K_d\in\compact(X)$ and a family of basic sets $(C^d_n)_{n\in\NN}$
such that:
\begin{itemize}
\item[-] $C^d_n\subseteq X_d$ for all $d\in E$ and $n\in\NN$
\item[-] $\hdm(K_d,C^d_n)\le 2^{-n}$ for all $n\in\NN$
\item[-] $K_d\subseteq X_d$ for all $d\in E$
\item[-] $\bigcup_{d\in E} K_d = K$
\end{itemize}
Moreover, if $X$ is a computable metric space such that for $y\in Q$, 
$\rho\in\QQ$ and $d\in D$ one can decide whether  
$\ball{\mu}{y}{\rho}\subseteq X_d$, then from $\varepsilon$ 
and $(C_n)_{n\in\NN}$ one can compute the families $(C^d_n)_{n\in\NN}$ as well as
a decision procedure for $E$.

\end{lemma}
\begin{proof}
Let $2 < m\in\NN$ such that $2^{-m} < \varepsilon/{8}$. 
Define, using decidabilty,
\[E := \{\, d\in D \mid 
 \exists y\in C_m \,\ball{\mu}{y}{\varepsilon/2}\subseteq X_d\,\}.\]
For $d\in E$ define $C^d_n\subseteq C_{m+n}$ by recursion on $n$:
\begin{align*}
&C^d_0 := \{\,y\in C_m\mid\ball{\mu}{y}{\varepsilon/2}\subseteq X_d\,\}\\
&C^d_{n+1} := \{\,y\in C_{m+n+1}\mid\exists z\in C^d_n\,\mu(y,z)\le 2^{1-(m+n)}\,\}
\end{align*}
\begin{claim}
\label{claim-split-cdn}
For all $d\in E$, $n\in\NN$ and  $z\in C^d_n$ there exists $y\in C^d_{n+1}$
such that $\mu(z,y)\le 2^{1-(n+m)}$.
\end{claim}
Proof: 
Let $z\in C^d_n\subseteq C_{m+n}$.
Let $x\in K$ such that $\mu(z,x)\le 2^{-(m+n)}$.
Let $y\in C_{m+n+1}$ such that $\mu(y,x)\le 2^{-(m+n+1)}$.
Then $\mu(z,y) \le 2^{1-(m+n)}$, hence $y\in C^d_{n+1}$.

\begin{claim}
\label{claim-split-cdn-cauchy}
For all $d\in E$ and $n\in\NN$, $C^d_n$ is a basic set and 
$\hdm(C^d_n,C^d_{n+1}) \le 2^{1-(m+n)}$.
\end{claim}
Proof: 
For the sets $C^d_n$ to be basic, it suffices to show their nonemptyness, since,
by definition, they are finite subsets of $Q$. Nonemptyness follows by induction
on $n$ using Claim~\ref{claim-split-cdn} and observing that $C^d_0$ is nonempty,
by definition.
The inequality $\hdm(C^d_n,C^d_{n+1}) \le 2^{1-(m+n)}$ follows from 
Claim~\ref{claim-split-cdn} and the definition of $C^d_{n+1}$.

\begin{claim}
\label{claim-split-cdn-subxd}
$C^d_n \subseteq X_d$, for all $d\in E$ and $n\in\NN$.
\end{claim}
Proof:
We show the stronger statement that 
$\cball{\mu}{C^d_n}{2^{2-(m+n)}}\subseteq X_d$, by induction on $n$. (Here, $\cball{\mu}{C^d_n}{2^{2-(m+n)}} = \bigcup\{\, \cball{\mu}{u}{2^{2-(m+n)}} \mid u \in C^d_n \,\}$.)
For the base case we have  
$\cball{\mu}{C^d_0}{2^{2-m}}\subseteq 
\ball{\mu}{C^d_0}{\varepsilon/2}\subseteq X_d$,
since $2^{2-m} < \varepsilon/2$.
For the step, let $y \in C^d_{n+1}$ and assume $\mu(y,x)\le 2^{2-(m+n+1)}$.
Let $z\in C^d_n$ with $\mu(y,z)\le 2^{1-(m+n)}$. Then 
$\mu(z,x) \le 2^{2-(m+n+1)}+2^{1-(m+n)} = 2^{2-(m+n)}$. Hence $x\in X_d$,
by induction hypothesis.
This ends the proof of Claim~\ref{claim-split-cdn-subxd}.

Since the space $\compact(X)$ is complete and, by 
Claim~\ref{claim-split-cdn-subxd}, for every $d\in E$, $(C^d_n)$ is a 
Cauchy sequence in $\compact(X)$, it has a limit, $K_d\in\compact(X)$.
\begin{claim}
\label{claim-split-cdnk}
$\hdm(K_d,C^d_n)<2^{-n}$.
\end{claim}
Proof: From Claim~\ref{claim-split-cdn-cauchy} it follows that 
$\hdm(K_d,C^d_n) \le 2^{2-(m+n)}$. But $2^{2-(n+m)} < 2^{-n}$, since $m > 2$.

\begin{claim}
\label{claim-split-kdsubk}
$K_d\subseteq K$.
\end{claim}
Proof: 
This follows from Lemma~\ref{lem-xink}~(b), since $C^d_n\subseteq C_{m+n}$.

\begin{claim}
\label{claim-split-kdsubxd}
$K_d\subseteq X_d$.
\end{claim}
Proof: 
Immediate, by Claim~\ref{claim-split-cdn-subxd} and Lemma~\ref{lem-xink}~(a).

\begin{claim}
\label{claim-split-union}
$\bigcup_{d\in E}K_d = K$.
\end{claim}
Proof: 
By Claim~\ref{claim-split-kdsubk}, it suffices to show that
$K \subseteq \bigcup_{d\in E}K_d$.
Let $x\in K$. Let $d\in D$ such that $\ball{\mu}{x}{\varepsilon}\subseteq X_d$.
Then $d\in E$, since there exists $y\in C_m$ with 
$\mu(x,y) < 2^{-m} < \varepsilon/4$, hence 
$\ball{\mu}{y}{\varepsilon/2}\subseteq X_d$.
We show that $x \in K_d$.
By Lemma~\ref{lem-xink}~(a), it suffices to show that for all $n\in\NN$ there
exists $y_n\in C^d_{m+n}$ such that $\mu(x,y_n) \le 2^{-n}$.
Since $\hdm(K,C^{m+n})<2^{-(m+n)}$, let $y_n\in C_{m+n}$ such that 
$\mu(x,y_n)<2^{-(n+m)}$.
We show that $y_n\in C^d_n$, by induction on $n$: $y_0 \in C_m$ and 
$\mu(x,y_0) < 2^{-m} < \varepsilon/4$, therefore 
$\ball{\mu}{y_0}{\varepsilon/2} \subseteq X_d$ (since 
$\varepsilon/2 + \varepsilon/4 < \varepsilon$). Hence $y_0\in C^d_0$.
For the induction step, we use the fact that $\mu(x,y_{n+1}) < 2^{-(n+m+1)}$, which
implies that $\mu(y_n,y_{n+1}) < 2^{1-(n+m)}$, hence $y_{n+1} \in C^d_{n+1}$.
This ends the proof of Claim~\ref{claim-split-union} and the proof of the Lemma.
\end{proof}

\begin{proof}[Proof of \fullref{thm-ak-sub-ck}]
 By Lemma~\ref{lem-p}, it suffices
to show that $(X,D,Q)$ has property (P). 
Therefore, let $K \in \compact(X)$ and $(C_n)_{n\in\NN}$ be a sequence of 
basic sets such that $\hdm(C_n,K) < 2^{-n}$ for all $n$. With
$X_d:= d[X]$, for $d \in D$, clearly, the hypotheses of Lemma~\ref{lem-split}
are satisfied. Hence we obtain $E\subseteq D$ and for each $d\in E$ a set 
$K'_d\in\compact(X)$ and a family of basic sets $(C'^d_n)_{n\in\NN}$
such that:
\begin{itemize}
\item[-] $C'^d_n\subseteq d[X]$ for all $d\in E$ and $n\in\NN$
\item[-] $\hdm(K'_d,C'^d_n)\le 2^{-n}$ for all $n\in\NN$
\item[-] $K'_d\subseteq d[X]$ for all $d\in E$
\item[-] $\bigcup_{d\in E} K'_d = K$
\end{itemize}
By Proposition~\ref{prop-choice-inv}, $d$ has a computable right inverse $d'$. Set $K^d := d'[K'_d]$. Then $K^d \in \compact(X)$. Since $d'$ has a computable modulus of continuity, we can, given $n \in \NN$, compute a number $k(n)$ so that for $x', x'' \in d[X]$, if $\mu(x', x'') < 2^{-k(n)}$ then $\mu(d'(x'), d'(x'')) < 2^{-n-1}$. It follows that $\hdm(K^d, d'[C'^d_{k(n)+1}]) < 2^{-n-1}$. By the computability of $d'$
we can moreover, for any $u \in C'^d_{k(n)+1}$, compute a basic element $v^n_u \in Q$ with $\mu(d'(u), v^n_u)) < 2^{-n-1}$.
Set $V_n := \set{v^n_u}{u \in C'^d_{k(n)+1}}$. Then $\hdm(d'[C'^d_{k(n)+1}], V_n) < 2^{-n-1}$. It follows that $\hdm(K^d, V_n) < 2^{-n}$. This completes the proof of Theorem~\ref{thm-ak-sub-ck}.
\end{proof}
 
 An important special case is if all digits have uniformly continuous inverses.
 
 \begin{definition}
 \label{def-unifinv}
 A digit space $(X, D,Q)$ is \emph{uniformly invertible} if for all $\theta > 0$ there exists (constructively) $\theta' > 0$ such that for all $d \in D$ and $x, y \in X$, if $\mu(d(x), d(y)) < \theta'$, then $\mu(x, y) < \theta$.
 \end{definition}
 
Note that any continuous injection of a compact space into a 
Hausdorff space is a homeomorphism on its image. Hence any injective
digit of a digit space has a uniformly continuous inverse. So the
only extra condition of uniform invertibility beyond injectivity is that
the inverse digits have \emph{effective} moduli of uniform continuity.

\begin{lemma}
\label{lem-unifinv-uniffib}
Let $(X, D,Q)$ be a uniformly invertible constructively dense computable digit space. Then $(X,D,Q)$ has approximable choice. 
\end{lemma}
\begin{proof}
Because of Proposition~\ref{prop-choice-inv} it remains to show that for every $d \in D$ and $u \in Q \cap d[X]$ we can compute a basic element $v \in Q$ such that $\mu(d^{-1}(u), v) < 2^{-n}$. For given $n \in \NN$ set $\theta_n := 2^{-n}$ and pick $\theta' \in \QQ_+$ according to uniform invertibility. Moreover, for given $u \in Q\cap d[X]$, use computable density to pick $v \in Q$ with $\mu(u, d(v)) < \theta'$. Then $\mu(d^{-1}(u), v) < \theta_n$.
 \end{proof}

\begin{corollary}
\label{cor-cauchy-coind}
Let $(X, D,Q)$  be a well-covering, decidable, uniformly invertible and constructively dense computable digit space. Then the following two statements hold:
\begin{enumerate}
\item\label{cor-cauchy-coind-1} $\cauchy = \coco_0$

\item\label{cor-cauchy-coind-2} $\cauchy_{\compact} = \coco_{\compact}$

\end{enumerate}
\end{corollary}

\section*{Acknowledgment}

The research leading to these results has received funding from the People Programme (Marie Curie Actions) of the European Union's Seventh Framework Programme FP7/2007-2013/ under REA grant agreement no.\ PIRSES-GA-2011-294962-COMPUTAL.

Research on this paper started on one of the many visits of the second author in Swansea. Main results were obtained while the first author was visiting the Department of Decision Sciences of the University of South Africa as part of the COMPUTAL project  and the second author was working in Pretoria as a Visiting Research Professor. 

Both authors are grateful to the Department of Decision Science in Pretoria for its hospitality and for having created such a splendid working atmosphere. The second author in addition also thanks the Computer Science Department of Swansea University for its great hospitality.

Thanks are also due to the anonymous referee for his careful reading of an earlier version of the manuscript.

\end{document}